\documentclass[aps,prb,reprint,amsmath,amssymb,superscriptaddress,longbibliography]{revtex4-2}

\usepackage{graphicx}
\usepackage{dcolumn}
\usepackage{bm}
\usepackage{amsmath}
\usepackage{amssymb}
\usepackage{xcolor}
\usepackage{xspace}
\usepackage{xfrac}
\usepackage{color,soul}
\usepackage{siunitx}
\usepackage{ulem}
\usepackage[colorlinks=true,urlcolor=blue,citecolor=blue,linkcolor=blue,bookmarks=false,pdfstartview={FitH}]{hyperref}
\newcommand{\bk}{{\bf k}}
\newcommand{\br}{{\bf r}}

\begin{document}

\author{Jinjing Yi}
\affiliation{Department of Physics and Astronomy, Center for Materials Theory, Rutgers University, Piscataway, NJ 08854, USA}
\author{Elio~J.\ K\"onig}
\affiliation{Max Planck Institute for Solid State Research, D-70569 Stuttgart, Germany}
\author{J. H. Pixley}
\affiliation{Department of Physics and Astronomy, Center for Materials Theory, Rutgers University, Piscataway, NJ 08854, USA}

\title{The low energy excitation spectrum of magic-angle semimetals}

\begin{abstract}
    We theoretically study the  excitation spectrum of a two-dimensional Dirac semimetal in the presence of an incommensurate potential. Such models have been shown to possess magic-angle critical points in the single particle wavefunctions, signalled by a momentum space delocalization of plane wave eigenstates and flat bands due to a vanishing Dirac cone velocity. {Using the kernel polynomial method,} we  compute the single particle Green's function to extract the nature of the single particle excitation energy, damping rate, and quasiparticle residue. As a result, we are able to clearly demonstrate the redistribution of spectral weight due to quasiperiodicity{-induced} downfolding {of} the Brillouin zone creating minibands with effective mini Brillouin zones that correspond to emergent superlattices. By computing the damping rate we show that the vanishing of the velocity and generation of finite density of states at the magic-angle transition coincides with the development of an imaginary part in the self energy and a {suppression} 
    of the quasiparticle residue that vanishes  in a power law like fashion. Observing these effects with ultracold atoms using momentum resolved radiofrequency spectroscopy is discussed.
\end{abstract}

\maketitle

\section{Introduction}

The ability to manipulate and control  the low-energy band structure in moir\'e heterostructures has enabled the observation of correlated insulating~\cite{cao2018correlated,chen2019signatures,ghiotto2021quantum,li2021continuous}, superconducting~\cite{CaoJarillo2018,LuEfetov2019,YankowitzDean2019}, and topological states~\cite{sharpe2019emergent,pixley2019ferromagnetism,serlin2020intrinsic,li2021quantum} despite starting from weakly interacting semiconducting materials. 
In a number of cases, the emergent many-body states are theoretically known to result from a dramatic renormalization of the band structure resulting in flat bands with a minibandwidth comparable to the Coulomb interaction scale~\cite{PhysRevLett.99.256802,TramblyMagaud2010,Morell2011,BistritzerMacDonald2011,PhysRevLett.121.026402,PhysRevLett.122.106405,PhysRevB.86.155449,PhysRevLett.125.257602,PhysRevB.103.205411,PhysRevB.103.205412,PhysRevB.103.205413,PhysRevB.103.205414,PhysRevB.103.205415,PhysRevB.103.205416}.
For Dirac semimetals, the bands become flat when the Dirac velocity vanishes due to a so-called ``magic-angle'' condition~\cite{BistritzerMacDonald2011}.
The magic-angle effect has been shown to be quite general and applicable to a broad class of nodal and topological band structures in the presence of a incommensurate tunneling or potential~\cite{PixleyGopalakrishnan2018,FuPixley2020,ChouPixley2020,PhysRevLett.126.103201,PhysRevA.100.053604,PhysRevLett.125.030504,PhysRevB.102.235126,PhysRevB.104.L041106,PhysRevB.103.155157,PhysRevX.11.021024,PhysRevB.104.L201113,volkov2020magic,tummuru2022josephson,lee2019theory,LeePixley2021,ledwith2021tb}.

\begin{figure}[b]
    \includegraphics[width=\linewidth]{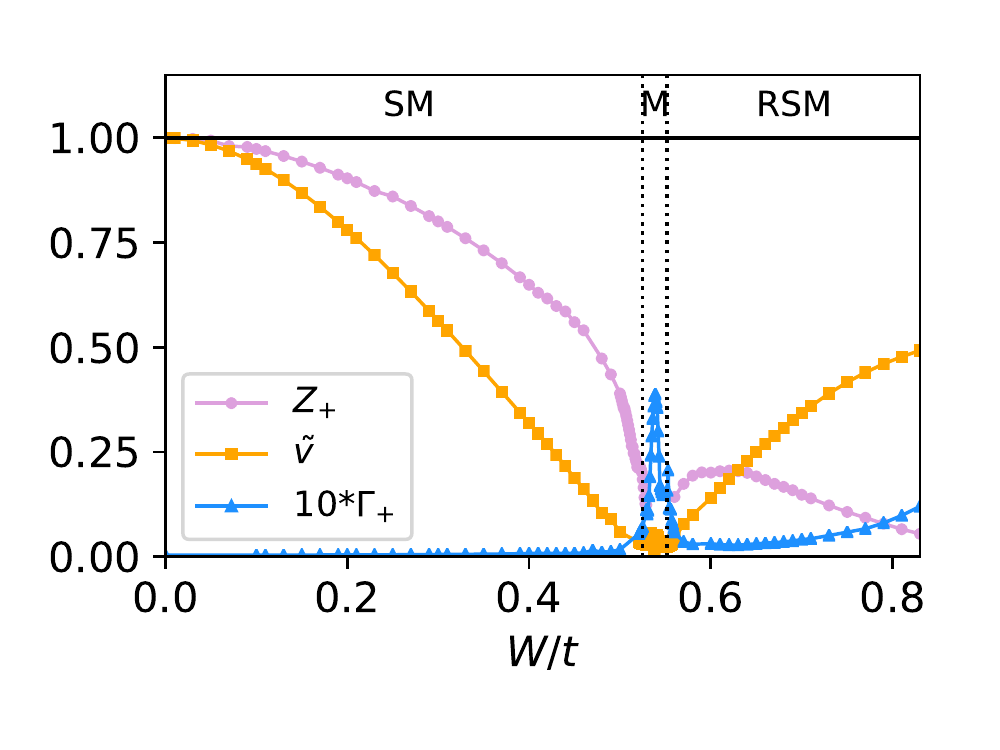}
    \caption{{\bf Phase diagram of the model defined in Eq.~\eqref{eqn:ham} at the Dirac node energy as a function of the quasperiodic potential strength $W/t$}:  The  Dirac velocity $\tilde v$ and the  quasiparticle residue $Z_+({\bf k}=0)$ are non-zero in the semimetal (SM) phase and the reentrant semimetal (RSM) phase at larger $W$. These regimes are separated by a metallic (M) phase with a finite density of states characterized by the imaginary part of the single particle self energy acquiring a non-zero value (multiplied by 10 for visual clarity). The dashed lines are the phase boundaries obtained from the density of states being non-zero and when the wavefunctions deloclalize in momentum space in Ref.~\cite{FuPixley2020}, demonstrating excellent agreement between the two approaches. These results were obtained {at the Dirac node ${\bf k}=0$}  with $L=144$ and  a kernel polynomial method expansion order $N_C=4096$. We note that the rise of $\Gamma$ at large $W$ is an $N_C$ artifact (see Fig.~\ref{fig:damping-Z}).}
    \label{fig:phase_diagram}
\end{figure}

Motivated by these developments, we here concentrate on the {\it momentum resolved  spectral features} of the magic-angle effect in two-dimensional (2D) Dirac semimetals. We specifically focus on the case when the moir\'e superlattice is incommensurate with the underlying microscopic lattice and are interested in  features beyond the continuum limit description.
A series of recent works revealed~\cite{PixleyGopalakrishnan2018,FuPixley2020,ChouPixley2020} 
the magic-angle {effect} as a
universal behavior that arises in a variety of models that possess two or three dimensional Dirac cones with an incommensurate modulation of the onsite scalar potential or  tunneling amplitude.
One major conclusion from this~\cite{FuPixley2020} and related work~\cite{PhysRevLett.125.030504} is that it is possible to realize the magic-angle phenomena without a twist, and instead this effect can be realized by smooth incommensurate perturbations.
In each case, the magic-angle condition drives the formation of flat bands and the concomitant formation of a finite density of states at the Dirac node energy. In the incommensurate limit, the magic-angle condition coincides with a single particle quantum phase transition between a semimetal and a metal where the wavefunctions Anderson delocalize in momentum space and the Dirac velocity vanishes in a universal fashion.

These findings have motivated a variety of quantum emulators of magic-angle graphene to be proposed \cite{FuPixley2020,ChouPixley2020}, particularly in the cold-atomic context\cite{GonzalezCirac2019,SalamonRakshit2020,PhysRevLett.126.103201}. In this regard, it is important to understand how to probe the magic-angle phenomena in a wide variety of systems. In this work, we focus on a  lattice model for magic-angle semimetals~\cite{FuPixley2020}, that can be realized using a quasi two-dimensional ultracold Fermi gas with artificial gauge fields~\cite{aidelsburger2015measuring,huang2016experimental,wu2016realization} and two incommensurate optical lattices~\cite{schreiber2015observation,PhysRevLett.122.110404}. 
We note that the considerations we focus on here are not limited to two dimensions;  the  realization of a three-dimensional Weyl semimetal using artificial  spin orbit coupling in ultracold gases~\cite{wang2021realization} raises the exciting prospect of observing magic-angle phenomena in higher dimensional semimetals~\cite{PixleyGopalakrishnan2018}.
In the context of ultracold gases, previous work has identified experimental signatures of the magic-angle transition through wavepacket dynamics~\cite{FuPixley2020} that can be measured with time-of-flight absorption imaging~\cite{foot2004atomic}, here we focus on the nature of the low-energy excitation spectrum that can be probed using momentum resolved radiofrequency spectroscopy~\cite{stewart2008using}. 

A natural starting point to describe the low energy excitations in moir\'e materials focuses on constructing continuum models, i.e. finding an effective superlattice in which to describe the downfolded and renormalized band structure. 
By construction, continuum models of moir\'e materials always have infinitely sharp and well defined  low energy quasiparticle excitations as the original microscopic lattice has been essentially coarse grained away. However, the question of whether or not these excitations survive and remain sharp when incommensurate effects of the twist are taken into account is a more subtle but physically relevant question. Along these lines, recent theoretical work has found that incommensurate effects at moderate ($\theta=9^{\circ}$ in the chiral limit)~\cite{FuPixley2020} and small twist angles ($\theta\approx 1.05^{\circ}$ i.e. the magic-angle)~\cite{gonccalves2021incommensurability} in twisted bilayer graphene can play a significant role when the velocity vanishes at the magic-angle through wavefunction multifractality in momentum space and subballistic transport. 
These incommensurate effects may also be relevant to  limits in the resolution of  recent angle resolved photoemission spectroscopy (ARPES) measurements of magic-angle twisted bilayer graphene~\cite{lisi2021observation,utama2021visualization}.

In this work we study the single-particle Green's function across the magic-angle quantum phase transition in a model of a 2D Dirac semimetal. In particular, we demonstrate that the flattening of the band is accompanied by a vanishing of the quasi particle residue, while the Anderson de-localization in momentum space is reflected in the onset of a finite decay rate in the momentum space Green's function that remains non-zero in the metallic phases of the model, {see Fig.~\ref{fig:phase_diagram}}. 
By examining the single-particle spectral function and the dispersion, our work shows how an incommensurate superlattice potential renormalizes the excitation spectrum inducing minibands with renormalized excitations described by an effective tight binding model on a series of enlarged length scales that correspond to a hierarchy of minibands.
Our work thus elucidates crucial features of Dirac electrons in the vicinity of the magic-angle and how to probe it within the spectral function that can be measured in both solid-state and ultracold atomic experimental set ups.

The remainder of the paper is structured as follows: In Sec.~\ref{sec:model}, we introduce the model under consideration and give details about the numerical approach. Sec.~\ref{sec:results} contains a summary of main results. We include a discussion, in which we compare to analytical approaches, Sec.~\ref{sec:Discussion}, before concluding with an outlook section~\ref{sec:Outlook}.

\section{Model and Approach}
\label{sec:model}

We consider a model of a 2D Dirac semimetal in the presence of an incommensurate potential
\begin{equation}
    \hat H=\hat H_0+\hat V.
    \label{eqn:ham}
\end{equation}
The 2D Dirac semimetal is realized by taking a tight binding model on the square lattice with spin orbit coupling
\begin{equation}
\hat{H}_{0} = \sum_{\mathbf{r},\mu=x,y}\left (\frac{it}{2}\hat\psi^{\dagger}_{\mathbf{r}}\sigma_{\mu}\hat\psi_{\mathbf{r}+a\hat\mu} + \textrm{h.c.}\right),
\label{eqn:ham0}
\end{equation}
where $\hat\psi_{\mathbf{r}}^{\dag}=(\hat c^{\dag}_{\mathbf{r},\uparrow},\hat c^{\dag}_{\mathbf{r},\downarrow})$ is a two component spinor of fermionic creation operators,  $\sigma_{\mu}$ are the Pauli matrices in the $\mu=x,y$ direction, $t$ is the hopping parameter for nearest neighbor lattice sites labeled by $\mathbf{r}$ with a lattice spacing $a$. {We remark that Eq.~\eqref{eqn:ham0} is equivalent to two copies of the $\pi$-flux model \cite{FuPixley2020}.}
Diagonalizing $H_0$ produces the dispersion,
\begin{equation}
    E^0_{\pm}(\bk)=\pm t \sqrt{\sin(ak_x)^2+\sin(ak_y)^2},
\end{equation}
that possess 4 Dirac cones $E^0_{\pm}(\bk)\approx \pm v|\bk|$ with $v=\pm ta$ at the 4 time reversal invariant points in the Brillouin zone $(0,0),(0,\pi/a),(\pi/a,0),(\pi/a,\pi/a)$.

The incommensurate potential is given by 
\begin{equation}
\hat{V} = W\sum_{\mathbf{r},\mu}\textrm{cos}(Qr_{\mu}+\phi_{\mu})\hat\psi^{\dagger}_{\mathbf{r}}\hat\psi_{\mathbf{r}},
\label{eqn:hamV}
\end{equation}
with a strength $W$, an incommensurate  wave-vector $Q=[2/(\sqrt{5}+1)]^2(2\pi/a)$, and the origin of the potential $\phi_{\mu}\in [0,2\pi)$ is randomly sampled with 30 realizations. 
To consider an incommensurate wavevector with periodic (or twisted) boundary conditions we take $Q$ to be a commensurate approximate $Q_n=(2\pi/a) F_{n-2}/F_n$ and fix the {linear} system size to be a Fibonacci number  $L=F_n$ such that $Q=\lim_{n\rightarrow \infty}Q_n$. 

\subsection{{Numerical Green's function approach}}

In the following 
we compute the retarded single particle Green's function for the model in Eqs.~\eqref{eqn:ham0} and \eqref{eqn:hamV} in frequency ($\omega$) and momentum ($\bk$) space
\begin{equation}
G_{\alpha\beta}(\bk,\omega) = \langle \bk, \alpha | \frac{1}{\omega + i \delta - \hat H}| \bk, \beta \rangle,
\label{eqn:G}
\end{equation}
where $\delta\rightarrow 0^+$, $\alpha,\beta$ denote the spin state $\uparrow,\downarrow$, and we have introduced momentum states
\begin{equation}
| \bk, \alpha \rangle = \frac{1}{L}\sum_{\br}e^{-i \br \cdot \bk}\hat\psi_{\br, \alpha}^{\dag}|0\rangle.
\end{equation}

 \begin{figure}[h]
\includegraphics[width=\linewidth, height=11cm]{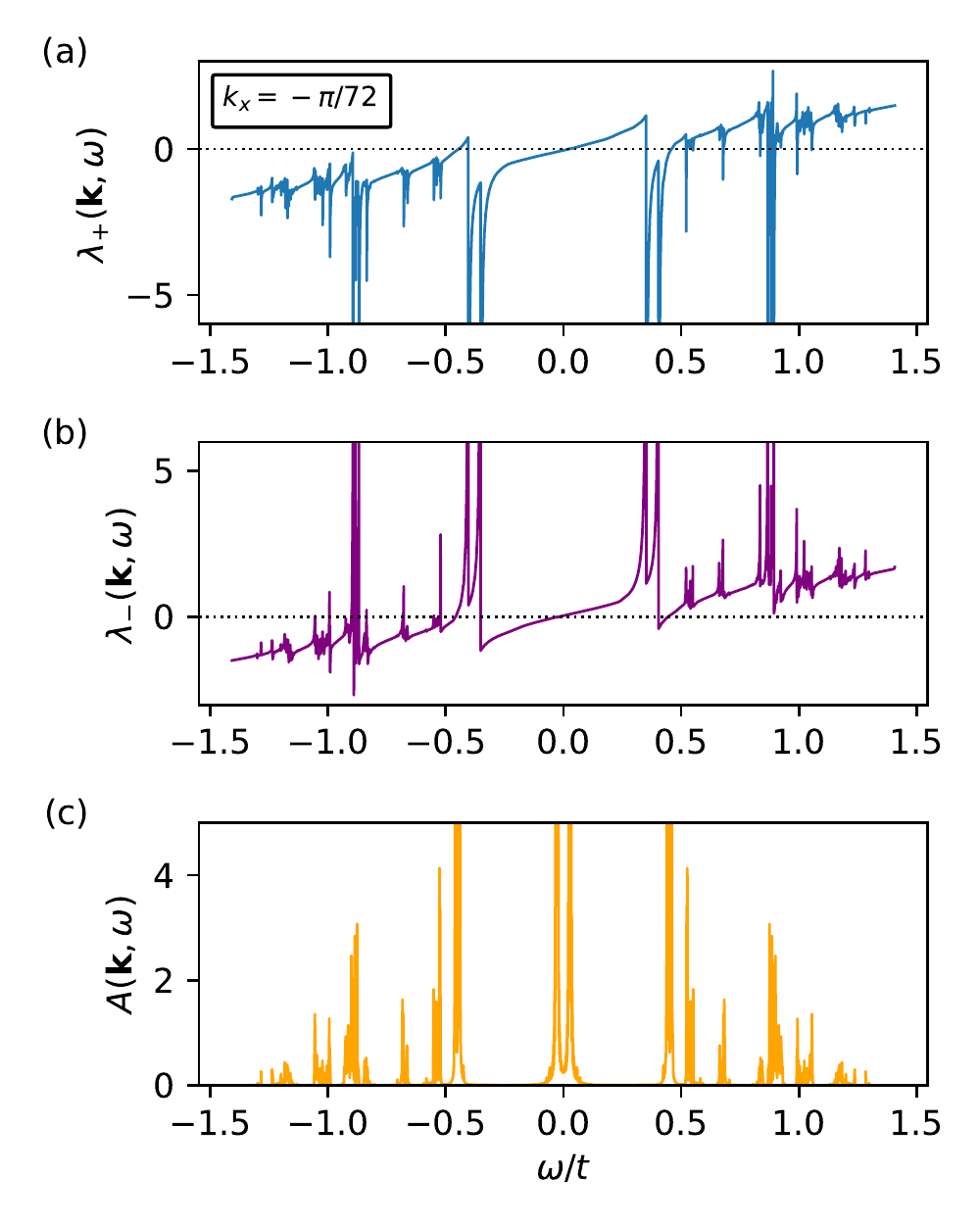}
\caption{
{\bf Extracting poles from the Green's function}: The top two panels (a) and (b) contain the real part of the two eigenvalues of the inverse Green's function $\lambda_{\pm}(\bk,\omega)\equiv\mathrm{Re}[1/G_{\pm}(\bk,\omega)]$ at $W=0.4t$ and the momentum ${\bf k}=(-\pi/72,0)$. The top plot is the real part of $\lambda_{+}({\bf k},\omega)$ as a function of $\omega/t$ and the middle plot is the real part of $\lambda_{-}({\bf k},\omega)$ as a function of $ \omega/t$. When $\lambda_{\pm}=0$ signals a pole in the Green's function allowing us to extract $E_{\pm}({\bf k})$. Note that the locations where $\lambda_{\pm}\rightarrow \infty$ represent band edges ({zeroes of the} Green's function) and we find additional excitations then occurring at higher frequency representing higher energy minibands due to the quasiperiodic potential downfolding and renormalizing the band structure. The bottom panel {c)} is the spectral function $A({\bf k},\omega)$ as a function of $\omega/t$, 
demonstrating that each crossing in $\lambda_{\pm}$ represents a sharp excitation creating a narrow Lorentzian in the spectral function. These results where obtained using KPM on a linear system size $L=144$ and expansion order $N_C=4096$ averaged over 30 samples. }
\label{fig:method}
\end{figure}

  We compute the Green's function perturbatively as well as in a numerically exact fashion using the kernel polynomial method (KPM)\cite{WeisseFehske2006}. Here we outline the KPM  approach for $G_{\alpha\beta}({\bf k},\omega)$ and the analysis we use to extract the physical content of the Green's function from the {behavior} 
 of its simple pole structure in the complex frequency plane. Expanding the Green's function in terms of Chebyshev polynomials, as described in Refs.~\cite{WeisseFehske2006,PixleyDasSarma2017}, we arrive at
\begin{eqnarray}
G_{\alpha\beta}(\bk,\omega) &=& \Big[-\frac{i}{a\sqrt{1-\tilde{\omega}^2}}\big(\mu_0(\bk, \alpha,\beta)g_0 
\nonumber
\\
&+& 
2\sum_{n=1}^{N_C-1} \mu_n(\bk, \alpha,\beta)g_n e^{-i n\arccos\tilde{\omega}}  \big)\Big],
\end{eqnarray}
where $g_n$ denotes the Kernel used to filter out Gibbs oscillations,  $a$ is the half-bandwidth, $b$ is half of the band asymmetry, $a=(E_{max}-E_{min})/(2-\epsilon)$, $b=(E_{max}+E_{min})/2$, where $E_{max}$ and $E_{min}$ are the maximum and minimum eigenvalues of the Hamiltonian and $\epsilon$ is a small parameter to avoid divergence, here $\epsilon=0.01$ \cite{WeisseFehske2006}, $N_C \gg 1$ is an integer denoting the truncated value of the series and $\tilde{\omega}=(\omega-b)/a$ is the rescaled energy. The coefficients of the expansion can be computed using sparse matrix-vector multiplication and are given by
\begin{equation}
\mu_n(\bk, \alpha,\beta) = \langle  \alpha, \bk | T_n(\tilde{H}) |\bk, \beta \rangle,
\end{equation}
where $\tilde{H}=(\hat{H}-b)/a$ is the rescaled Hamiltonian that has eigenvalues that lie within $[-1,1]$.
The Lorentz kernel $g_n$~\cite{WeisseFehske2006} is used to preserve the analytic properties of the Green function, which broadens each Dirac-delta function in the spin-resolved spectral function $A_{\alpha\beta}(\bk,\omega)=-\mathrm{Im}G_{\alpha\beta}(\bk,\omega)/\pi$ into a Lorentzian~\cite{WeisseFehske2006} of width $\Lambda D/N_C$ (for a bandwidth $D$), and $\Lambda$ controls both the width of the Lorentzian and the strength of Gibbs oscillations due to truncating the series.  The full spectral function is given by 

\begin{equation}
A(\bk,\omega) = - \frac{1}{2\pi}\text{Im} \lbrace\text{Tr}[G(\bk, \omega)]\rbrace=\frac{1}{2}\sum_{\alpha=\uparrow,\downarrow}A_{\alpha\alpha}(\bk,\omega).
\label{eqn:Akw}
\end{equation} 
Here, we work with $\lambda=0.5$ so that we can accurately compute the intrinsic broadening.
In the following we focus on a linear system size $L=144$ where the rounding due to the KPM expansion order dominates over finite size rounding.

\subsection{Extracting the excitation spectrum}

To extract the physical content from the full Green's function in Eq.~\eqref{eqn:G} we diagonalize the 2x2 matrix $G_{\alpha\beta}(\bk,\omega)$ for each $\bk$ and $\omega$ corresponding to two complex eigenvalues $G_{\pm}(\bk,\omega)$. To understand the pole structure we focus on the real part of the inverse of these eigenvalues denoted
$\lambda_{\pm}(\bk,\omega)=\mathrm{Re}[1/G_{\pm}(\bk,\omega)]$, see Fig.~\ref{fig:method} (a), (b). An excitation of the spectrum is realized as a pole in the Green's function, thus $\lambda_{\pm}{(\bk, \omega)}$ will go through zero at the location of the pole, defining the dispersion $\omega = E_{\pm}(\bk)$, and producing a large amount of spectral weight in the (effective) Lorentzians in the spectral function $A(\bk,\omega)$ as shown in Fig.~\ref{fig:method} (c). To extract $Z_{\pm}(\bk)${, $\tilde v$ and $\gamma_\pm(\bk)$} we expand the eigenvalues near the poles that yields 
\begin{equation}
    G_\pm(\bk, \omega) \simeq \frac{Z_\pm(\bk)}{\omega - E_\pm(\bk) + i \gamma_\pm(\bk)}. \label{eq:GFinverseMainText}
\end{equation}
To estimate $Z^{-1}_{\pm}(\bk)$ we fit $\lambda_{\pm}(\bk,\omega)$ to a linear function of $\omega-E_{\pm}(\bk)$  for $\omega$ in close proximity to $E_{\pm}(\bk)$ to extract the slope. Moreover, we obtain $\tilde v$ from the small $\bk$ expansion of $E_\pm(\bk)\approx \pm\tilde{v}|{\bf k}|$ and $\gamma_\pm(\bk) =-Z(\bk) \text{Im}[G_{\pm}(\bk, \omega = E(\bk))^{-1}]$.

In relation to the above approximation of $G$ near a simple pole, it is instructive to consider the Dyson equation, 
\begin{equation}
    G(\bk,\omega)^{-1}=G_0(\bk,\omega)^{-1}-\Sigma(\bk,\omega), \label{eqn:Dyson}
\end{equation}
to understand the effect of the incommensurate potential. Firstly, in the Dirac semimetal phase  of the model we expect that the potential only generates a real part of the self energy at low energies $\Sigma_{\pm}(\bk,\omega)=\Sigma'_{\pm}(\bk,\omega)$ that renormalizes the velocity of the Dirac cone $\tilde v$ and the quasiparticle residue $Z(\bk)$ whereas the imaginary part $\Sigma''_{\pm}(\bk,0)=0$ in the thermodynamic limit. Thus, it is natural to expect that  the metallic phase of the model will be signalled by the generation of a non-zero $\mathrm{Im} \Sigma_{\pm}(0,0)$, which in turn can lead to a finite density of states at the Dirac node. In full analogy to Anderson's original definition\cite{Anderson1958}, the onset of $\mathrm{Im} \Sigma_{\pm}(0,0) \neq 0$ delineates the delocalization transition in momentum space.

To separate out the effects of the quasiparticle residue and the generation of a non-zero $\mathrm{Im} \Sigma_{\pm}(\bk,\omega)$ we  study these two contributions separately. To do so we focus on an effective damping rate on-shell that is not suppressed by the residue, namely
\begin{equation}
    \Gamma_{\pm}(\bk)\equiv \mathrm{Im}\Sigma(\bk,\omega=E_{\pm}(\bk))
    \label{eqn:imsigma}
\end{equation}
and are able to track the phases of the model by focusing on $\Gamma(0)$ as shown in Fig.~\ref{fig:phase_diagram}. We note that the quasiparticle damping rate is then given by $\gamma_{\pm}(\bk)=Z_{\pm}(\bk)\Gamma_{\pm}(\bk)$.

Due to the quasiperiodic potential the low energy spectrum is strongly renormalized. This appears in the band structure as effectively downfolding the Brillouin zone to create a whole new set of excitations that live within a renormalized miniband. As shown in Fig.~\ref{fig:method}, in addition to the original Dirac excitation expected near zero energy we see several new poles appear in the Green's function. Moreover, we also find zero's of the real part of the Green's function that signify the edge of a miniband. Note that it is important not to artificially identify such a divergence with a pole as the figure makes it look. We now turn to understanding this low energy excitation spectrum in detail.

\subsection{{Perturbation theory}}

As the Dirac semimetal is perturbatively stable, we can substatiante this perspective as well as understand the structure of the Green function in the semimetal phase by applying perturbation theory. The self-energy to leading order in the potential is~\cite{FuPixley2020} 
\begin{equation}
    \Sigma(\bk, \omega) = \sum_{ \mu = 1}^d \sum_{\xi = \pm} \frac{W^2}{4} G_0(\bk + \xi Q \hat e_\mu, \omega), \label{eq:PerturbationTheory}
\end{equation}
where $d = 2$ and $\hat e_\mu \in \lbrace (1,0)^T, (0,1)^T \rbrace$ for $d = 2$. The solution of this equation leads to~\cite{FuPixley2020} a  quasiparticle residue
\begin{align}
    Z_{\pm}(\bk =0) &= \frac{1}{1 + 4\alpha^2}, \label{eq:Zpert}
\end{align}
and a renormalized Dirac velocity 
\begin{align}
    \tilde v &=  \frac{ 1 - 2(1- \cos(Q))\alpha^2 }{1+4\alpha^2}
    \label{eqn:vpert}
\end{align}
in terms of a dimensionsless coupling $\alpha=W/[2t\sin(Q)]$.
While we find that the residue remains non-zero for any $\alpha$, $\tilde v$ contains a magic-angle condition defined by when the renormalized velocity vanishes that occurs perturbatively at $\alpha_{{\rm c, pert.}}^2=1/[2(1-\cos(Q)]$. We stress that the existence of a magic-angle extends beyond perturbation theory as demonstrated in several models using numerics~\cite{PixleyGopalakrishnan2018,FuPixley2020,ChouPixley2020}, though the location shifts depending on symmetries and form of the model.
 Thus,  perturbatively  the quasiparticle residue remains finite when the model passes through the magic-angle. As we will show below, in the incommensurate limit the residue vanishes in a power law fashion at the magic-angle.

\begin{figure*}[htb!]
\includegraphics[width=0.99\linewidth]{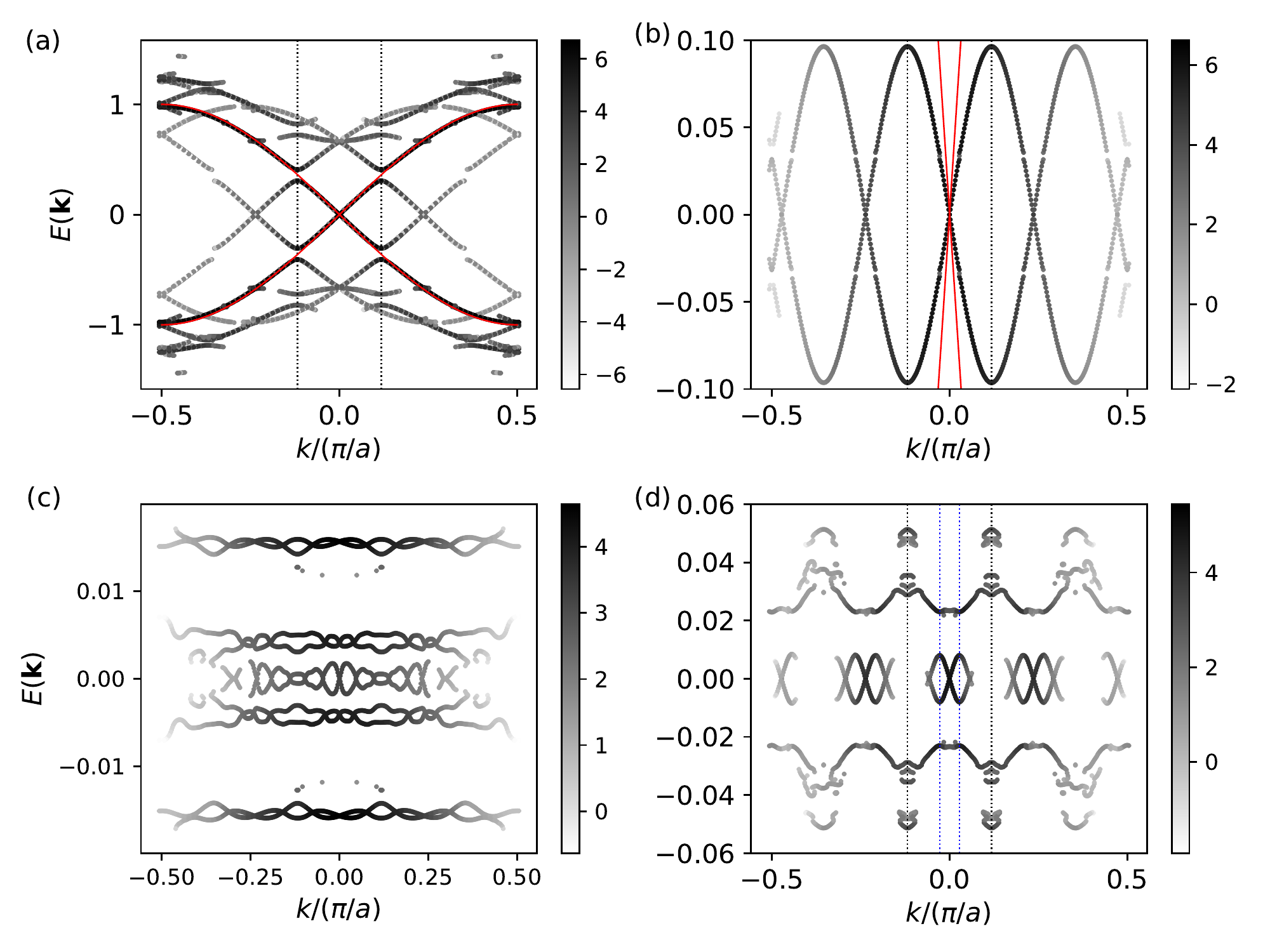}

\caption{{\bf Low energy dispersion relation} extracted from the Green's function in each phase of the model, in the semimetal (a) $W=0.1t$ and (b) $W=0.4t$; the metal (c) $W=0.54t$; and the reentrant semimetal (d) $W=0.6t$. These figures were obtained from the KPM with $N_C$=4096 for $L=144$. The red line in (a) and (b) is the dispersion for $W=0$. The {gray} 
scale denotes the value of the spectral weight of each excitation (in particular $\log A({\bf k},E_{\pm}({\bf k})$ is shown). 
In (a) we show the full energy range to  see the start of the formation of the first mininband. In (b) we see the first miniband has formed as the dispersion is repeating outside the first mini Brillouin zone (BZ) of size $G_{\mathrm{MB}_1}$, marked with vertical black dashed lines. {The onset of the second BZ downfolding is manifested by small discontinuities at $k$ corresponding to $G_{{\rm MB}_2}\simeq 0.06\pi/a$.} In the metallic phase (c), the bands have become {extremely} flat {(note the rescaled ordinate)} and several new minibands have formed. We expect the finite number of minibands here is attributed to the finite system size we have considered. In the reentrant semimetal phase the second miniband has formed of extent 
$G_{\mathrm{MB}_2}$ (marked with blue vertical dashed lines). The miniband dispersions in (b) and (d) are folded into their respective BZs in Fig.~\ref{fig:folded_dispersion}. 
}
\label{fig:dispersion}
\end{figure*}

\section{Results}
\label{sec:results}

Using the method described above, we 
extract the
excitation spectrum and renormalized spectral weight  from the Green's function. While previous work has inferred the nature of the excitation spectrum of the model defined in Eq.~\eqref{eqn:ham} from the density of states and the response of eigenvalues to twisted boundary conditions, here, we extract this excitation spectrum directly from numerically computing   $G(\bk,\omega)$ on a finite system size averaged over 30 samples of different phases of the quasiperiodic potential.  Due to the symmetries of the model after we average it is sufficient to consider the momentum along the $k_x$ direction, i.e. without loss of generality we only focus on $\bk=(k,0)$.

For the given choice of $Qa/2\pi=[2/(\sqrt{5}+1)]^2$ and focusing on the range of $W/t\lesssim 1$, as shown in Fig.~\ref{fig:phase_diagram} the model has three relevant phases of interest, a Dirac semimetal phase, a metallic phase, and a reentrant Dirac semimetal phase. The excitations only remain sharp  in the semimetal phase   albeit with a renormalized Dirac velocity that decreases on approach to the transition but importantly the imaginary part of the self energy vanishes in the thermodynamic limit. This is also consistent with the non-zero value of the quasiparticle residue at the Dirac node $Z_{\pm}(\bk=0)$. In contrast, the metallic phase develops a finite $\Gamma(\bk=0)$ concomitantly with a finite density of states and the quasiparticle residue vanishes in a power law fashion on approach to the metallic phase (in the metallic regime we find it is no longer possible to accurately estimate the residue). In the metallic phase the low energy excitations are no longer sharp and the nature of this damping is investigated further in Sec.~\ref{sec:ZandGamma}. 

\begin{figure}[b!]
\includegraphics[width=0.99\linewidth]{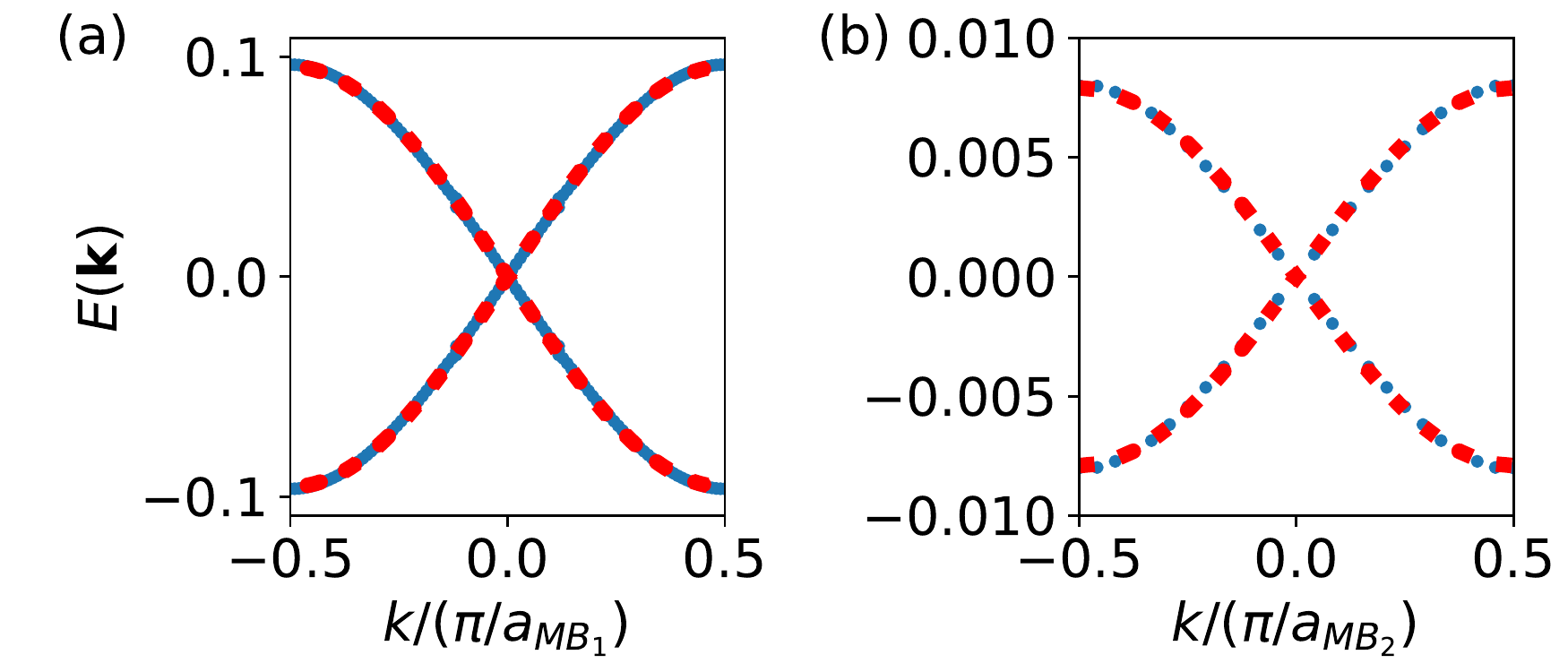} 
\caption{{ \bf Folded band structure into the first and second mBZs}: (a) The band structure for $W=0.4t$ in the semimetal phase from the top panel of Fig.~\ref{fig:dispersion} folded into the first mBZ.
(b) The band structure for $W=0.6 t$ folded into the second mBZ.
We fit each folded band structure (shown as red dashed lines) to the form for tunneling on a square moir\'e lattice in Eq.~\eqref{eqn:mini BZ} with a greatly renormalized tunneling amplitude.
}
\label{fig:folded_dispersion}
\end{figure}

\begin{figure*}[htb!]

\includegraphics[width=\linewidth]{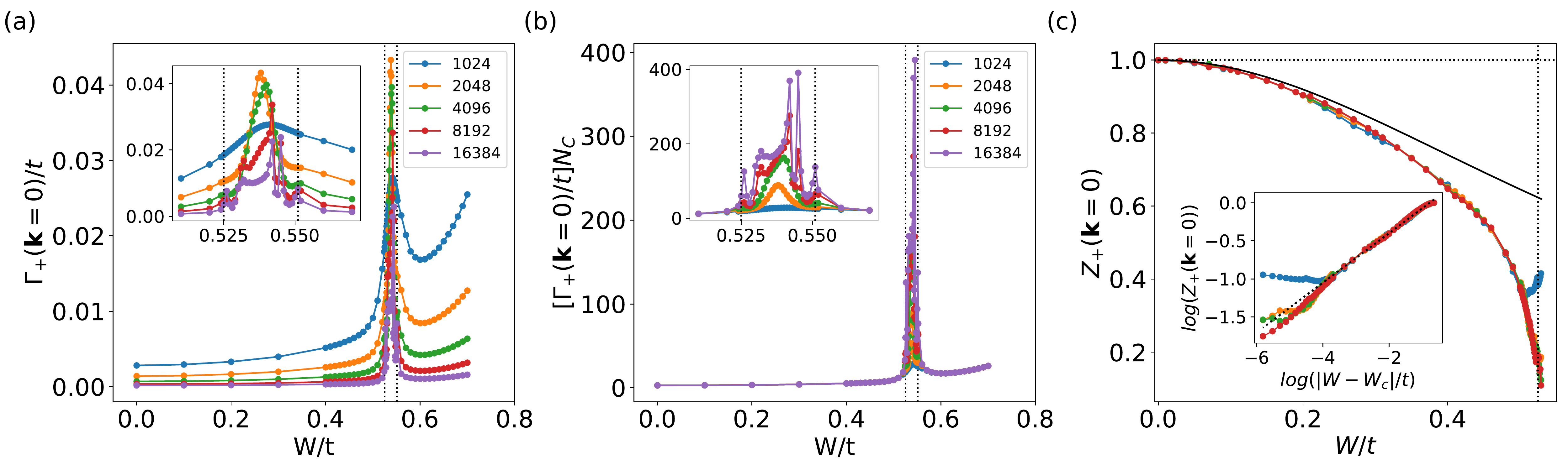}

\caption{{\bf Scaling of the damping rate and residue}: The left panel (a) shows the damping rate $\Gamma_{+}$ and the middle panel (b) shows the scaled damping rate, $N_c\Gamma_{+}$, demonstrating that damping in each semimetal phase behaves  
like $\Gamma_{+}\sim 1/N_C$ and the damping is solely extrinsic due to the KPM broadening in these regimes. 
As shown in the insets as $N_C$ gets larger, we can see that  $\Gamma_{+}$ is sharpening up while developing additional structure that survives the large $N_C$ limit. The right panel {c)} shows the $N_C$ dependence of the residue for $W\leq 0.52$,  We can see the residue starts from 1 at $W=0.0$ and decreases {and eventually} vanishes in a power law fashion  {$Z_{+} \sim |W-W_c|^{y}$} with $y=0.33\pm0.04$ near $W_c=0.525$~\cite{FuPixley2020} via the fit shown as a dashed line on approach to the magic-angle transition (inset). The solid black line is the residue from perturbation theory, Eq.~\eqref{eq:Zpert}.} 

\label{fig:damping-Z}
\end{figure*}

\subsection{Dispersion and spectral weight}

The dispersion is shown in Fig.~\ref{fig:dispersion} in each of these phases along with the corresponding spectral weight $A(\bk,\omega)$ in Eq.~\eqref{eqn:Akw} at $\omega=E_{\pm}(\bk)$ (depicted via a grey scale). 
First, we start from a weak potential strength of $W=0.1t$ in the semimetal phase shown in Fig.~\ref{fig:dispersion} (a). We find that a miniband has formed due to a gap appearing at finite energy that separates the Dirac semimetal excitations at low energy from the rest of the spectrum at higher energies.
For this potential the velocity of the Dirac excitations is only weakly renoramlized (the $W=0$ dispersion is shown in red).
Further into the semimetal phase as shown in Fig.~\ref{fig:dispersion}(b) we see that the Dirac cone still remains stable but with a  renormalized velocity. Interestingly, we find a new Brillioun zone has emerged as the dispersion is now periodic with a periodicity corresponding to a new inverse length scale that is not  simply $\pi/a$ (or a rational multiple of it). This mini Brillouin zone (mBZ) can be understood by considering the leading perturbative process that scatters the Dirac cones in the problem (for the model in Eqs.~\eqref{eqn:ham} they are separated by $\pi$ in momentum space). This produces a new mBZ of size $G_{\mathrm{MB}_1}=(\pi-Q)/a = \pi (\sqrt{5} - 2)/a\approx 0.24 \pi/a$, 
(with reciprocal lattice vectors ${\bf G}_{\mathrm{MB}_1}/G_{\mathrm{MB}_1}=\hat x,\hat y$),
which is precisely where the dispersion becomes periodic~\cite{FuPixley2020}. 
Intuitively, this can be interpreted as the excitations now living on the moir\'e superlattice of size $a_{\mathrm{MB}_1}={2\pi}/G_{\mathrm{MB}_1} = 2a /(\sqrt{5} - 2) \approx 8.5 a$
as we explore in more detail below.
At the same time the spectral weight is redistributed due to the downfolding of the BZ. We find the spectral weight is maximal in the first mBZ and the higher order zones of the first mBZ ($|k|>G_{\mathrm{MB}_1}$) have decreasing spectral weight. We can also see the second mBZ forming where the higher zones are missing excitations.

The properties of the reentrant semimetal phase are similar to the original semimetal regime but as we can see in Fig.~\ref{fig:dispersion}(d) a second mBZ has opened up with a much smaller size. We also see additional higher order second mBZs with smaller spectral weigth appearing. This arises due to fourth order scattering processess producing a second mBZ of size $G_{\mathrm{MB}_2}=-(3\pi-4Q)/a = (9 - 4 \sqrt{5}) \pi/a\approx 0.056 \pi/a$. Thus, the excitations of the second mBZ live on a much larger length scale $a_{\mathrm{MB}_2}=2\pi/G_{\mathrm{MB}_2}= 2a/(9-4\sqrt{5})\approx 35.9a$. 

These results clearly demonstrate that the semimetal phases of the model have well defined low energy Dirac excitations that live within effective mBZs. 
To examine this in more detail in Fig.~\ref{fig:folded_dispersion} we have folded the spectrum for $W=0.4t$ into the first mBZ and folded the spectrum into the second mBZ for $W=0.6t$. We find that the dispersion in each mBZ $E_{\pm}^{\mathrm{MB}}(\bk)$  follows that of the original model with a renormalized hopping strength $t_{\mathrm{MB}}$ and enlarged lattice spacing $a_{\mathrm{MB}}$ corresponding to the emergent superlattice, namely
\begin{equation}
    E_{\pm}^{\mathrm{MB}}(\bk)=\pm t_{\mathrm{MB}}\sqrt{\sin(k_xa_{\mathrm{MB}})^2 + \sin(k_ya_{\mathrm{MB}})^2}
    \label{eqn:mini BZ}
\end{equation} as shown in Fig.~\ref{fig:folded_dispersion} (as dashed red lines). Thus, in the semimetal phases of the model we can describe the low energy physics 
as a tight-binding model 
on an emergent superlattice, whose size depends on the overall strength of the incommensurate potential. Which miniband is occuring at low energy depends on the vicinity to the magic-angle transition and it has been argued to become an indefinite process in the thermodynamic limit at the magic-angle~\cite{FuPixley2020}.

In contrast to the semimetal regimes, in the metallic phase Fig.~\ref{fig:dispersion} (c) we find several flat bands with a greatly renormalized energy scale (comparing the vertical scale of the panels of Fig.~\ref{fig:dispersion}). This is consistent with the vanishing Dirac cone velocity (e.g. see Fig.~\ref{fig:phase_diagram}). 
{As the Dirac velocity has dropped substantially, the system is very susceptible to the incommensuration and new physics emerges close to charge neutrality.}
Thus, this regime requires more physical insight then the dispersion alone, and therefore we turn to the nature of excitations at the Dirac node.

\subsection{Residue and damping at the Dirac node}

\label{sec:ZandGamma}

To characterize the nature of the Dirac point we compute the quasiparticle residue, $Z_{\pm}(\bk=0)$, the Dirac velocity $\tilde v$ from the low energy excitations $E_{\pm}(\bk)\approx  \pm \tilde v |\bk|$ and a measure of the damping rate $\Gamma_{\pm}(\bk=0)$ that reflects  a finite imaginary part of the self energy rate that is insensitive to changes in the residue.
We note that at $\bk=0$ the $\pm$ eigenvalues of the Green function are degenerate so we just take the $+$ value here. While previous work has identified the magic-angle transition  through measuring $\tilde v$ from the low energy scaling of the density of states, we here compute $\tilde v$ directly from the extracted excitation spectrum.

The presence of sharp and coherent Dirac excitations in the semimetal phases is exemplified through a finite quasiparticle residue and velocity as well as a vanishing damping rate in the thermodynamic limit. As shown in Fig.~\ref{fig:phase_diagram} the quasiparticle residue is regular in the semimetal phase and so the damping rate is essentially determined by $\Gamma_{\pm}(\bk)$, see below Eq.~\eqref{eqn:imsigma}. We therefore focus on $\Gamma_{\pm}(\bk=0)$ as shown in Fig.~\ref{fig:damping-Z} (a).  In the semimetal phase of the model, we expect that there is no intrinsic quasiparticle lifetime and these exictations are therefore infinitely long-lived in the thermodynamic limit. Thus in our finite size KPM simulations the damping rate should be an extrinsic effect due to the method, and  we expect that $\Gamma_{\pm}(0)\sim 1/N_C$, in each semimetal phase in excellent agreement with the collapsed data in  Fig.~\ref{fig:damping-Z} (b). In the metallic phase of the model we find that the self energy has developed an intrinsic non-zero imaginary part that does not vanish for increasing $N_C$ but  is instead developing additional structure for increasing  $N_C$, see the insets of Fig.~\ref{fig:damping-Z} (a) and (b).

At this point it is instructive to compare this result to real space Anderson delocalization based on the locator expansion~\cite{Anderson1958}. Here, delocalization is observed by when the local self energy acquires an imaginary part demonstrating the presence of resonant states that destabalize the localized phase. Similarly, here we find that the self energy at the Dirac node ${\bf k}=0$ becomes non-zero in the metallic phase of the model, where the single particle wavefunctions have been shown to Anderson delocalize in momentum space~\cite{FuPixley2020}. The excellent agreement between the phase boundaries in Fig.~\ref{fig:phase_diagram} solidifies this connection.

As previously discussed, the extracted velocity decreases and vanishes on approach to the metallic phase from either semimetal regime as shown in Fig.~\ref{fig:phase_diagram}. These numerical results are consistent with the perturbative expectation in Eq.~\eqref{eqn:vpert} of a vanishing velocity $\tilde v\sim |W-W_c|$ at the transition. The presence of sharp and coherent Dirac excitations in the semimetal phases of the model is also clearly demonstrated by the finiteness of the quasiparticle residue at the Dirac point. As seen in Fig.~\ref{fig:damping-Z} (c) we find $Z_{\pm}(\bk=0)$ is finite in the semimetal phases of the model  
(in the metallic phase we find it is not well defined as $Z_{\pm}(\bk=0)$  changes sign). Near the magic-angle transition at $W_c=0.525\pm0.005$ for the $Q$ considered here~\cite{FuPixley2020}, our data  is consistent with $Z$ vanishing like a power law, namely
\begin{equation}
    Z_{\pm}(\bk=0)\sim |W-W_c|^y.
\end{equation}
A fit this power-law form is shown in Fig.~\ref{fig:damping-Z} (c) (inset), which yields an exponent 
$y = 0.33\pm 0.04$. This exponent suggests a non-zero anomalous dimension $\eta$ at the magic-angle transition. If we assume conventional scaling relations, consistent with a continuous transition with a diverging correlation length $\xi\sim |W-W_c|^{-\nu}$ we have $\mathrm{Re}G({\bf k},\omega=0)\sim 1/k^{1-\eta}$, which we can relate to the Green function in the semimetal regime to obtain $Z(\bk=0)\sim \xi^{-\eta}\sim|W-W_c|^{\eta\nu}$ 
where we find $\eta \nu \approx 0.33$.

The data in the close proximity to $W_c$ becomes strongly $N_C$ dependent, see Fig.~\ref{fig:damping-Z} (c) inset, which represents a  finite  energy resolution effect similar to a finite size effect~\cite{PhysRevB.94.121107}. The power law vanishing of $Z_{\pm}(0)$  is also affected by the finite system size and the intrinsic rounding introduced by our method to extract $Z$ based on a fit to the low energy form near the pole (see Sec.~\ref{sec:model}). We thus attribute the finiteness of the residue at the transition  to these  extrinsic effects of our computational approach.  Note that if we instead compute $Z_{\pm}(\bk=0)$ perturbatively we find that it does not vanish at the magic-angle and to second order in the potential we find $Z_{\pm}(\bk=0) {\vert_{W = W_c}= 1/[1+2/(1-\cos(Q))]}$, see Eq.~\eqref{eq:Zpert}, suggesting that the vanishing of the residue arises due to strong coupling incommensurate effects at the transition.

\section{Discussion}
\label{sec:Discussion}

The momentum resolved spectral features we have studied clearly show the presence of the magic-angle phenomena that appears as a vanishing quasiparticle residue and the onset of a non-zero quasiparticle lifetime, in addition to the already expected vanishing velocity. The model we have focused on in Eqs.~\eqref{eqn:ham0},\eqref{eqn:hamV}, can be realized in an ultracold Fermi gas using artificial gauge fields. The spectral functions we have studied can be directly measured in this set up using momentum resolved radiofrequency spectroscopy as has been done in the Fermi Hubbard model\cite{stewart2008using}. The width of the spectral peak at zero momentum can be used to measure the quasiparticle damping rate, while the amplitude can be related to the quasiparticle residue. The presence of a harmonic trap will round out the singular behavior of the magic-angle transition into a cross over, making a box trap~\cite{PhysRevLett.110.200406} the ideal setting to explore this transition.

As we have seen in this manuscript, perturbation theory provides an excellent platform to interpret a number of the observed effects we see such as the formation of minibands with excitations on the moir\'e length scale and a vanishing velocity. However, perturbatively the  quasiparticle residue does not vanish  and the quasiparticle damping rate remains zero, in stark contrast to what we find in this incommensurate limit. Thus, it is a natural question to ask how do we construct a theoretical description for the phenomena associated with the magic-angle that we have found. Along these lines, a natural approach is extending the perturbative expression of the self energy to a self-consistent framework. Here, we summarize perhaps one of the simplest self-consistent theories, 
and why it eventually is not capable to accurately account for the observed phenomena and leave the development of a proper analytical treatment for the future.

{Returning to the Dyson equation presented in Eq.~\eqref{eqn:Dyson}, the simplest self-consistent generalization to Eq.~\eqref{eq:PerturbationTheory} is to replace $G_0(\bk, \omega)$ on the right hand side by $G(\bk, \omega)$ given in  Eq.~\eqref{eqn:Dyson}
\begin{equation}
    \Sigma(\bk, \omega) = \sum_{ \mu = 1}^d \sum_{\xi = \pm} \frac{W^2}{4} G(\bk + \xi Q \hat e_\mu, \omega). \label{eq:SelfConsistent}
\end{equation}
This is the quasiperiodic variant of the standard self-consistent Born-approximation in the theory of disordered systems. 
}
Note that Eq.~\eqref{eq:SelfConsistent} can be seen as a locator expansion in momentum space and resembles the method used by Sokoloff~\cite{Sokoloff1980} to study higher-dimensional Aubry-Andr\'e models. As each operator-insertion of $\hat V$ amounts to a hopping by $\pm Q \hat e_{x,y}$, 
$\hat V$ induced contribution to the self-energy can be represented graphically as closed momentum space trajectories consistent of hops at distance $Q$. Out of all possible contributions, the self-consistent approximation, Eq.~\eqref{eq:SelfConsistent}, keeps only
semiclassical trajectories which originate from $\bk$ (i.e. loops are neglected).

\begin{figure}
\includegraphics[width=\linewidth]{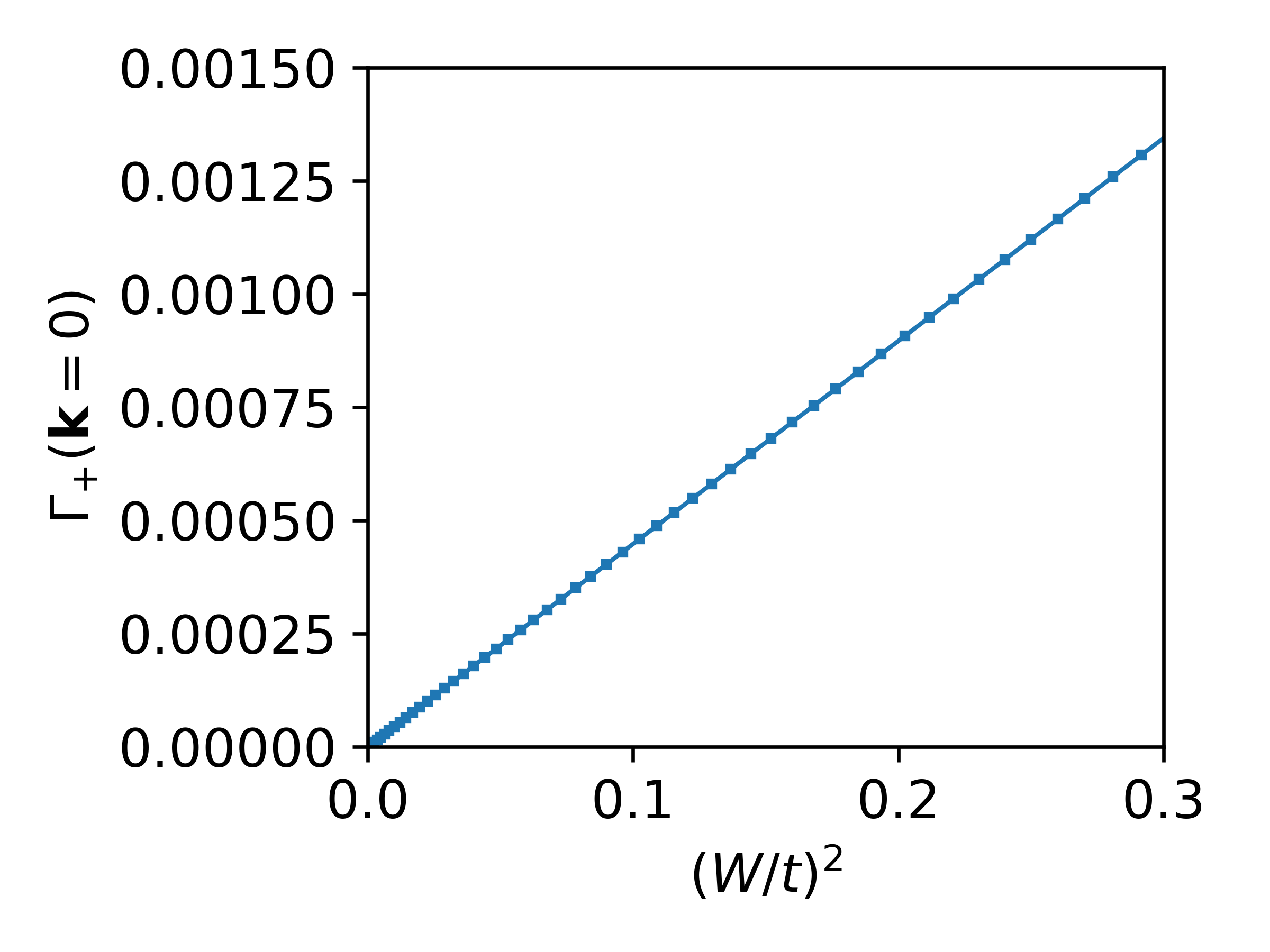} 
\caption{{Solution of the imaginary part of the self energy using the self-consistent equation~\eqref{eq:SelfConsistent} for}, $L=144$, $\delta=0.00001$, $Q=\pi {110}/144$, where $\delta$ acts as the small imaginary part of the initial self energy. }
\label{fig:SelfConsistent}
\end{figure}

We have solved the self-consistent equations numerically, Fig.~\ref{fig:SelfConsistent}, but unfortunately we find the self-consistent treatment leads to a finite decay rate even at infinitesimal strength of incommensuration (for a discussion see Appendix \ref{app:SelfConsistent}).
We therefore leave the study of an analytical theory for the eigenstate quantum phase transition to the future. We speculate, that an appropriate large-$N$ generalization of our model, for which Eq.~\eqref{eq:SelfConsistent} is the controlled saddle point equation, will display
the critical state 
in a $1/N$ vicinity of $W/t= 0$.

\section{Conclusion and Outlook}
\label{sec:Outlook}

In this work we have studied the single particle Green's function across magic-angle phase transitions in incommensurate two-dimensional Dirac semimetals. We find that the stable semimetal phase consists of renormalized low energy Dirac excitations on an emergent moir\'e lattice that changes as the strength of the incommensurate potential is varied. This results in a decreased velocity and quasiparticle residue. Interestingly, we find that in the incommensurate limit at the magic-angle transition the quasiparticle residue vanishes and the imaginary part of the single-particle self-energy becomes non-zero, signalling a non-zero density of states at the Dirac node.

It will be very interesting to explore related effects in models of twisted bilayer graphene that incorporate incommensurate effects. Also, the approach we have used can also be helpful in analyzing ARPES data on twisted van der Waals heterostructures~\cite{lisi2021observation,utama2021visualization,PhysRevB.103.235146} as our results suggest that near the magic-angle the excitations will have an {\it intrinsic broadening} due to incommensurate effects of the twist that will affect the resolution of the low-energy flat bands. Keeping in mind that the states at the Dirac node are certainly special, we conclude by highlighting that similar phenomena might also be of relevance in semimetallic twisttronics systems at large twist angles~\cite{Ahn2018,ParkLee2019} or at small twist angles away from the Dirac energy~\cite{AttigRosch2021}. Beyond twisttronics, the comprehensive study of low-energy excitations at eigenstate quantum phase transitions in higher dimensional quasiperiodic~\cite{AubryAndre1980} and quasicrystalline systems\cite{DevakulHuse2017,SzaboSchndeider2020} remains thus a formidable task for the future.

\acknowledgments

It is a pleasure to thank Y.Z. Chou,  Yixing Fu, and  J. H. Wilson for discussions and related previous work. 
This work was partially supported by Grant No. 2020264 from the United States-Israel Binational Science Foundation (BSF), Jerusalem, Israel (J.Y. J.H.P.), the Air Force Office of Scientific Research under Grant No.~FA9550-20-1-0136 (J.H.P.) and the Alfred P. Sloan Foundation through a Sloan Research Fellowship (J.H.P.).
E.J.K. and J.H.P. acknowledge hospitality by the Aspen Center for Physics, where part of this work was completed and which is
supported by National Science Foundation grant PHY1607611.
The authors acknowledge the following research computing resources: the Beowulf cluster at the Department of Physics and Astronomy of Rutgers University, and the Amarel cluster from the Office of Advanced Research Computing (OARC) at Rutgers, The State University of New Jersey \cite{amarel}. 

\appendix

\section{Self-consistent theory and large-N approach}
\label{app:SelfConsistent}

This appendix provides additional details for the  effective self-consistent Eq.~\eqref{eq:SelfConsistent} of the main text, including details for the behavior observed numerically in Fig.~\ref{fig:SelfConsistent} and an appropriate large-$N$ limit. 

\subsection{Explanation of numerical observations}

In Fig.~\ref{fig:SelfConsistent} we present the numerical solution of the self-consistent equation, which displays a damping reate $\Gamma(\bk = 0) \propto W^2/t$. This can be explained as follows: First, note that a non-zero $\text{Im}\Sigma(\bk, \omega)$ for any given $\bk$ implies non-zero $\text{Im}\Sigma(\bk', \omega)$ at another momentum $\bk'$ at some order $N(\bk, \bk')$ of perturbation theory in $W$. Essentially this reflects Mott's classic arguments~\cite{Mott1967} according to which either all states at a given energy $\omega$ are localized in $\bk$-space ($\text{Im}\Sigma(\bk, \omega) = 0, \forall \bk$), or all states are delocalized ($\text{Im}\Sigma(\bk, \omega) \neq 0, \forall \bk$). Second, note that $\text{Im}\Sigma(\bk, \omega) $ is negative semidefinite by causality. Third, taking the integral of Eq.~\eqref{eq:SelfConsistent} leads to
\begin{equation}
\int_{\rm BZ} \frac{d^d k}{(2\pi)^d}  \text{Im}\text{tr}[\Sigma(\bk, \omega)] = - \pi d \frac{W^2}{4} \nu(\omega), \label{eq:app:IntegratedSigma}
\end{equation}
where $\nu(\omega) = -\int_{\rm BZ} {d^d k} \text{Im}G(\bk, \omega)/(2^d \pi^{d+1})$ is the density of states. 

We thus find that the integral over a negative semi-definite function is finite if $\nu(\omega) \neq 0$, which implies that the function is non-zero. By Mott's argument it even implies $\text{Im}\Sigma(\bk, \omega) < 0, \forall \bk$. Thus, Eqs.~\eqref{eq:SelfConsistent}, \eqref{eq:app:IntegratedSigma} rigorously imply a non-zero decay rate at least for all momentum eigenstates at non-zero energies. Finally, we discuss the limit $\omega = 0$. Considering that a finite decay rate implies non-zero $\nu(\omega)$ even at $\omega = 0$ and assuming a continuity of the self-energy both as a function of $W$ and $\omega$, this explains the numerical finding, $\text{Im}\Sigma(\bk=0, \omega = 0) \neq 0$.

\subsection{Large-N treatment}

{What is a model which features Eq.~\eqref{eq:SelfConsistent} as a controlled mean field equation? We found a large-$N$ generalization of Eq.~\eqref{eqn:hamV}, in which an additional quantum number $a,b = 1, \dots N$ of fermions is introduced and the Hamiltonian is replaced by 
\begin{subequations}
\begin{eqnarray}
    \hat H_0 &=& \sum_{\mathbf{r}, \mu = x,y, a} \frac{it}{2} \psi^\dagger_{\mathbf{r}, a} \sigma_\mu \psi_{\mathbf r + \hat \mu, a} + H.c.  \\
\hat V &=& \sum_{ab, \mu}\frac{W_{a b}^\mu}{2} \sum_{\mathbf{r}} e^{i Q r_\mu} \psi^\dagger_{\mathbf r,a} \psi_{\mathbf r, b} + H.c.
\end{eqnarray}
\label{eq:LargeNHam}
\end{subequations}
where $W_{a b}^\mu$ are complex random $N \times N$ matrices with only non-zero moment $\langle W_{ab}^{\mu,*}W_{a'b'}^{\mu'} \rangle = \delta_{aa'} \delta_{bb'}\delta_{\mu \mu'} W^2/N$. This ensures that the total number of $e^{i Q r_\mu}$ insertions is balanced by the same number of $e^{-i Q r_\mu}$ insertions, and translational symmetry is restored. It also ensures average flavor symmetry (i.e. $\Sigma_{aa'} \propto \delta_{aa'}$). In this model, the physics at each site in real space is thus essentially given by random matrix theory at large $N \gg 1$ where the leading diagrams for the self-energy are given by rainbow diagrams. The resummation of such diagrams leads to Eq.~\eqref{eq:SelfConsistent} of the main text. As explained there, it is equivalent to keeping semiclassical trajectories of hopping in momentum space. 

It is important to emphasize that, despite the random matrix character, Eq.~\eqref{eq:LargeNHam} does not correspond to a disorder model. In a given realization, the matrices $W_{ab}^\mu$ are the same for all sites, i.e., in contrast to Hamiltonians describing physical disorder, the number of random variables does not diverge in the thermodynamic limit. It is also important to observe that Eq.~\eqref{eq:LargeNHam} is not the same as Eq.~$\eqref{eqn:ham}$ when $N \rightarrow 1$, since both amplitude and phase of $W$ fluctuate. At the same time,
in analogy to random matrix theory, the rainbow diagrams in our model are responsible for a finite density of states. }

\bibliography{main}

\begin{thebibliography}{69}%
\makeatletter
\providecommand \@ifxundefined [1]{%
 \@ifx{#1\undefined}
}%
\providecommand \@ifnum [1]{%
 \ifnum #1\expandafter \@firstoftwo
 \else \expandafter \@secondoftwo
 \fi
}%
\providecommand \@ifx [1]{%
 \ifx #1\expandafter \@firstoftwo
 \else \expandafter \@secondoftwo
 \fi
}%
\providecommand \natexlab [1]{#1}%
\providecommand \enquote  [1]{``#1''}%
\providecommand \bibnamefont  [1]{#1}%
\providecommand \bibfnamefont [1]{#1}%
\providecommand \citenamefont [1]{#1}%
\providecommand \href@noop [0]{\@secondoftwo}%
\providecommand \href [0]{\begingroup \@sanitize@url \@href}%
\providecommand \@href[1]{\@@startlink{#1}\@@href}%
\providecommand \@@href[1]{\endgroup#1\@@endlink}%
\providecommand \@sanitize@url [0]{\catcode `\\12\catcode `\$12\catcode
  `\&12\catcode `\#12\catcode `\^12\catcode `\_12\catcode `\%12\relax}%
\providecommand \@@startlink[1]{}%
\providecommand \@@endlink[0]{}%
\providecommand \url  [0]{\begingroup\@sanitize@url \@url }%
\providecommand \@url [1]{\endgroup\@href {#1}{\urlprefix }}%
\providecommand \urlprefix  [0]{URL }%
\providecommand \Eprint [0]{\href }%
\providecommand \doibase [0]{https://doi.org/}%
\providecommand \selectlanguage [0]{\@gobble}%
\providecommand \bibinfo  [0]{\@secondoftwo}%
\providecommand \bibfield  [0]{\@secondoftwo}%
\providecommand \translation [1]{[#1]}%
\providecommand \BibitemOpen [0]{}%
\providecommand \bibitemStop [0]{}%
\providecommand \bibitemNoStop [0]{.\EOS\space}%
\providecommand \EOS [0]{\spacefactor3000\relax}%
\providecommand \BibitemShut  [1]{\csname bibitem#1\endcsname}%
\let\auto@bib@innerbib\@empty
%</preamble>
\bibitem [{\citenamefont {Cao}\ \emph {et~al.}(2018{\natexlab{a}})\citenamefont
  {Cao}, \citenamefont {Fatemi}, \citenamefont {Demir}, \citenamefont {Fang},
  \citenamefont {Tomarken}, \citenamefont {Luo}, \citenamefont
  {Sanchez-Yamagishi}, \citenamefont {Watanabe}, \citenamefont {Taniguchi},
  \citenamefont {Kaxiras} \emph {et~al.}}]{cao2018correlated}%
  \BibitemOpen
  \bibfield  {author} {\bibinfo {author} {\bibfnamefont {Y.}~\bibnamefont
  {Cao}}, \bibinfo {author} {\bibfnamefont {V.}~\bibnamefont {Fatemi}},
  \bibinfo {author} {\bibfnamefont {A.}~\bibnamefont {Demir}}, \bibinfo
  {author} {\bibfnamefont {S.}~\bibnamefont {Fang}}, \bibinfo {author}
  {\bibfnamefont {S.~L.}\ \bibnamefont {Tomarken}}, \bibinfo {author}
  {\bibfnamefont {J.~Y.}\ \bibnamefont {Luo}}, \bibinfo {author} {\bibfnamefont
  {J.~D.}\ \bibnamefont {Sanchez-Yamagishi}}, \bibinfo {author} {\bibfnamefont
  {K.}~\bibnamefont {Watanabe}}, \bibinfo {author} {\bibfnamefont
  {T.}~\bibnamefont {Taniguchi}}, \bibinfo {author} {\bibfnamefont
  {E.}~\bibnamefont {Kaxiras}}, \emph {et~al.},\ }\bibfield  {title} {\bibinfo
  {title} {Correlated insulator behaviour at half-filling in magic-angle
  graphene superlattices},\ }\href@noop {} {\bibfield  {journal} {\bibinfo
  {journal} {Nature}\ }\textbf {\bibinfo {volume} {556}},\ \bibinfo {pages}
  {80} (\bibinfo {year} {2018}{\natexlab{a}})}\BibitemShut {NoStop}%
\bibitem [{\citenamefont {Chen}\ \emph {et~al.}(2019)\citenamefont {Chen},
  \citenamefont {Sharpe}, \citenamefont {Gallagher}, \citenamefont {Rosen},
  \citenamefont {Fox}, \citenamefont {Jiang}, \citenamefont {Lyu},
  \citenamefont {Li}, \citenamefont {Watanabe}, \citenamefont {Taniguchi} \emph
  {et~al.}}]{chen2019signatures}%
  \BibitemOpen
  \bibfield  {author} {\bibinfo {author} {\bibfnamefont {G.}~\bibnamefont
  {Chen}}, \bibinfo {author} {\bibfnamefont {A.~L.}\ \bibnamefont {Sharpe}},
  \bibinfo {author} {\bibfnamefont {P.}~\bibnamefont {Gallagher}}, \bibinfo
  {author} {\bibfnamefont {I.~T.}\ \bibnamefont {Rosen}}, \bibinfo {author}
  {\bibfnamefont {E.~J.}\ \bibnamefont {Fox}}, \bibinfo {author} {\bibfnamefont
  {L.}~\bibnamefont {Jiang}}, \bibinfo {author} {\bibfnamefont
  {B.}~\bibnamefont {Lyu}}, \bibinfo {author} {\bibfnamefont {H.}~\bibnamefont
  {Li}}, \bibinfo {author} {\bibfnamefont {K.}~\bibnamefont {Watanabe}},
  \bibinfo {author} {\bibfnamefont {T.}~\bibnamefont {Taniguchi}}, \emph
  {et~al.},\ }\bibfield  {title} {\bibinfo {title} {Signatures of tunable
  superconductivity in a trilayer graphene moir{\'e} superlattice},\
  }\href@noop {} {\bibfield  {journal} {\bibinfo  {journal} {Nature}\ }\textbf
  {\bibinfo {volume} {572}},\ \bibinfo {pages} {215} (\bibinfo {year}
  {2019})}\BibitemShut {NoStop}%
\bibitem [{\citenamefont {Ghiotto}\ \emph {et~al.}(2021)\citenamefont
  {Ghiotto}, \citenamefont {Shih}, \citenamefont {Pereira}, \citenamefont
  {Rhodes}, \citenamefont {Kim}, \citenamefont {Zang}, \citenamefont {Millis},
  \citenamefont {Watanabe}, \citenamefont {Taniguchi}, \citenamefont {Hone}
  \emph {et~al.}}]{ghiotto2021quantum}%
  \BibitemOpen
  \bibfield  {author} {\bibinfo {author} {\bibfnamefont {A.}~\bibnamefont
  {Ghiotto}}, \bibinfo {author} {\bibfnamefont {E.-M.}\ \bibnamefont {Shih}},
  \bibinfo {author} {\bibfnamefont {G.~S.}\ \bibnamefont {Pereira}}, \bibinfo
  {author} {\bibfnamefont {D.~A.}\ \bibnamefont {Rhodes}}, \bibinfo {author}
  {\bibfnamefont {B.}~\bibnamefont {Kim}}, \bibinfo {author} {\bibfnamefont
  {J.}~\bibnamefont {Zang}}, \bibinfo {author} {\bibfnamefont {A.~J.}\
  \bibnamefont {Millis}}, \bibinfo {author} {\bibfnamefont {K.}~\bibnamefont
  {Watanabe}}, \bibinfo {author} {\bibfnamefont {T.}~\bibnamefont {Taniguchi}},
  \bibinfo {author} {\bibfnamefont {J.~C.}\ \bibnamefont {Hone}}, \emph
  {et~al.},\ }\bibfield  {title} {\bibinfo {title} {Quantum criticality in
  twisted transition metal dichalcogenides},\ }\href@noop {} {\bibfield
  {journal} {\bibinfo  {journal} {Nature}\ }\textbf {\bibinfo {volume} {597}},\
  \bibinfo {pages} {345} (\bibinfo {year} {2021})}\BibitemShut {NoStop}%
\bibitem [{\citenamefont {Li}\ \emph {et~al.}(2021{\natexlab{a}})\citenamefont
  {Li}, \citenamefont {Jiang}, \citenamefont {Li}, \citenamefont {Zhang},
  \citenamefont {Kang}, \citenamefont {Zhu}, \citenamefont {Watanabe},
  \citenamefont {Taniguchi}, \citenamefont {Chowdhury}, \citenamefont {Fu}
  \emph {et~al.}}]{li2021continuous}%
  \BibitemOpen
  \bibfield  {author} {\bibinfo {author} {\bibfnamefont {T.}~\bibnamefont
  {Li}}, \bibinfo {author} {\bibfnamefont {S.}~\bibnamefont {Jiang}}, \bibinfo
  {author} {\bibfnamefont {L.}~\bibnamefont {Li}}, \bibinfo {author}
  {\bibfnamefont {Y.}~\bibnamefont {Zhang}}, \bibinfo {author} {\bibfnamefont
  {K.}~\bibnamefont {Kang}}, \bibinfo {author} {\bibfnamefont {J.}~\bibnamefont
  {Zhu}}, \bibinfo {author} {\bibfnamefont {K.}~\bibnamefont {Watanabe}},
  \bibinfo {author} {\bibfnamefont {T.}~\bibnamefont {Taniguchi}}, \bibinfo
  {author} {\bibfnamefont {D.}~\bibnamefont {Chowdhury}}, \bibinfo {author}
  {\bibfnamefont {L.}~\bibnamefont {Fu}}, \emph {et~al.},\ }\bibfield  {title}
  {\bibinfo {title} {Continuous mott transition in semiconductor moir{\'e}
  superlattices},\ }\href@noop {} {\bibfield  {journal} {\bibinfo  {journal}
  {Nature}\ }\textbf {\bibinfo {volume} {597}},\ \bibinfo {pages} {350}
  (\bibinfo {year} {2021}{\natexlab{a}})}\BibitemShut {NoStop}%
\bibitem [{\citenamefont {Cao}\ \emph {et~al.}(2018{\natexlab{b}})\citenamefont
  {Cao}, \citenamefont {Fatemi}, \citenamefont {Fang}, \citenamefont
  {Watanabe}, \citenamefont {Taniguchi}, \citenamefont {Kaxiras},\ and\
  \citenamefont {Jarillo-Herrero}}]{CaoJarillo2018}%
  \BibitemOpen
  \bibfield  {author} {\bibinfo {author} {\bibfnamefont {Y.}~\bibnamefont
  {Cao}}, \bibinfo {author} {\bibfnamefont {V.}~\bibnamefont {Fatemi}},
  \bibinfo {author} {\bibfnamefont {S.}~\bibnamefont {Fang}}, \bibinfo {author}
  {\bibfnamefont {K.}~\bibnamefont {Watanabe}}, \bibinfo {author}
  {\bibfnamefont {T.}~\bibnamefont {Taniguchi}}, \bibinfo {author}
  {\bibfnamefont {E.}~\bibnamefont {Kaxiras}},\ and\ \bibinfo {author}
  {\bibfnamefont {P.}~\bibnamefont {Jarillo-Herrero}},\ }\bibfield  {title}
  {\bibinfo {title} {Unconventional superconductivity in magic-angle graphene
  superlattices},\ }\href@noop {} {\bibfield  {journal} {\bibinfo  {journal}
  {Nature}\ }\textbf {\bibinfo {volume} {556}},\ \bibinfo {pages} {43}
  (\bibinfo {year} {2018}{\natexlab{b}})}\BibitemShut {NoStop}%
\bibitem [{\citenamefont {Lu}\ \emph {et~al.}(2019)\citenamefont {Lu},
  \citenamefont {Stepanov}, \citenamefont {Yang}, \citenamefont {Xie},
  \citenamefont {Aamir}, \citenamefont {Das}, \citenamefont {Urgell},
  \citenamefont {Watanabe}, \citenamefont {Taniguchi}, \citenamefont {Zhang}
  \emph {et~al.}}]{LuEfetov2019}%
  \BibitemOpen
  \bibfield  {author} {\bibinfo {author} {\bibfnamefont {X.}~\bibnamefont
  {Lu}}, \bibinfo {author} {\bibfnamefont {P.}~\bibnamefont {Stepanov}},
  \bibinfo {author} {\bibfnamefont {W.}~\bibnamefont {Yang}}, \bibinfo {author}
  {\bibfnamefont {M.}~\bibnamefont {Xie}}, \bibinfo {author} {\bibfnamefont
  {M.~A.}\ \bibnamefont {Aamir}}, \bibinfo {author} {\bibfnamefont
  {I.}~\bibnamefont {Das}}, \bibinfo {author} {\bibfnamefont {C.}~\bibnamefont
  {Urgell}}, \bibinfo {author} {\bibfnamefont {K.}~\bibnamefont {Watanabe}},
  \bibinfo {author} {\bibfnamefont {T.}~\bibnamefont {Taniguchi}}, \bibinfo
  {author} {\bibfnamefont {G.}~\bibnamefont {Zhang}}, \emph {et~al.},\
  }\bibfield  {title} {\bibinfo {title} {Superconductors, orbital magnets and
  correlated states in magic-angle bilayer graphene},\ }\href@noop {}
  {\bibfield  {journal} {\bibinfo  {journal} {Nature}\ }\textbf {\bibinfo
  {volume} {574}},\ \bibinfo {pages} {653} (\bibinfo {year}
  {2019})}\BibitemShut {NoStop}%
\bibitem [{\citenamefont {Yankowitz}\ \emph {et~al.}(2019)\citenamefont
  {Yankowitz}, \citenamefont {Chen}, \citenamefont {Polshyn}, \citenamefont
  {Zhang}, \citenamefont {Watanabe}, \citenamefont {Taniguchi}, \citenamefont
  {Graf}, \citenamefont {Young},\ and\ \citenamefont
  {Dean}}]{YankowitzDean2019}%
  \BibitemOpen
  \bibfield  {author} {\bibinfo {author} {\bibfnamefont {M.}~\bibnamefont
  {Yankowitz}}, \bibinfo {author} {\bibfnamefont {S.}~\bibnamefont {Chen}},
  \bibinfo {author} {\bibfnamefont {H.}~\bibnamefont {Polshyn}}, \bibinfo
  {author} {\bibfnamefont {Y.}~\bibnamefont {Zhang}}, \bibinfo {author}
  {\bibfnamefont {K.}~\bibnamefont {Watanabe}}, \bibinfo {author}
  {\bibfnamefont {T.}~\bibnamefont {Taniguchi}}, \bibinfo {author}
  {\bibfnamefont {D.}~\bibnamefont {Graf}}, \bibinfo {author} {\bibfnamefont
  {A.~F.}\ \bibnamefont {Young}},\ and\ \bibinfo {author} {\bibfnamefont
  {C.~R.}\ \bibnamefont {Dean}},\ }\bibfield  {title} {\bibinfo {title} {Tuning
  superconductivity in twisted bilayer graphene},\ }\href@noop {} {\bibfield
  {journal} {\bibinfo  {journal} {Science}\ }\textbf {\bibinfo {volume}
  {363}},\ \bibinfo {pages} {1059} (\bibinfo {year} {2019})}\BibitemShut
  {NoStop}%
\bibitem [{\citenamefont {Sharpe}\ \emph {et~al.}(2019)\citenamefont {Sharpe},
  \citenamefont {Fox}, \citenamefont {Barnard}, \citenamefont {Finney},
  \citenamefont {Watanabe}, \citenamefont {Taniguchi}, \citenamefont
  {Kastner},\ and\ \citenamefont {Goldhaber-Gordon}}]{sharpe2019emergent}%
  \BibitemOpen
  \bibfield  {author} {\bibinfo {author} {\bibfnamefont {A.~L.}\ \bibnamefont
  {Sharpe}}, \bibinfo {author} {\bibfnamefont {E.~J.}\ \bibnamefont {Fox}},
  \bibinfo {author} {\bibfnamefont {A.~W.}\ \bibnamefont {Barnard}}, \bibinfo
  {author} {\bibfnamefont {J.}~\bibnamefont {Finney}}, \bibinfo {author}
  {\bibfnamefont {K.}~\bibnamefont {Watanabe}}, \bibinfo {author}
  {\bibfnamefont {T.}~\bibnamefont {Taniguchi}}, \bibinfo {author}
  {\bibfnamefont {M.}~\bibnamefont {Kastner}},\ and\ \bibinfo {author}
  {\bibfnamefont {D.}~\bibnamefont {Goldhaber-Gordon}},\ }\bibfield  {title}
  {\bibinfo {title} {Emergent ferromagnetism near three-quarters filling in
  twisted bilayer graphene},\ }\href@noop {} {\bibfield  {journal} {\bibinfo
  {journal} {Science}\ }\textbf {\bibinfo {volume} {365}},\ \bibinfo {pages}
  {605} (\bibinfo {year} {2019})}\BibitemShut {NoStop}%
\bibitem [{\citenamefont {Pixley}\ and\ \citenamefont
  {Andrei}(2019)}]{pixley2019ferromagnetism}%
  \BibitemOpen
  \bibfield  {author} {\bibinfo {author} {\bibfnamefont {J.~H.}\ \bibnamefont
  {Pixley}}\ and\ \bibinfo {author} {\bibfnamefont {E.~Y.}\ \bibnamefont
  {Andrei}},\ }\bibfield  {title} {\bibinfo {title} {Ferromagnetism in
  magic-angle graphene},\ }\href@noop {} {\bibfield  {journal} {\bibinfo
  {journal} {Science}\ }\textbf {\bibinfo {volume} {365}},\ \bibinfo {pages}
  {543} (\bibinfo {year} {2019})}\BibitemShut {NoStop}%
\bibitem [{\citenamefont {Serlin}\ \emph {et~al.}(2020)\citenamefont {Serlin},
  \citenamefont {Tschirhart}, \citenamefont {Polshyn}, \citenamefont {Zhang},
  \citenamefont {Zhu}, \citenamefont {Watanabe}, \citenamefont {Taniguchi},
  \citenamefont {Balents},\ and\ \citenamefont {Young}}]{serlin2020intrinsic}%
  \BibitemOpen
  \bibfield  {author} {\bibinfo {author} {\bibfnamefont {M.}~\bibnamefont
  {Serlin}}, \bibinfo {author} {\bibfnamefont {C.}~\bibnamefont {Tschirhart}},
  \bibinfo {author} {\bibfnamefont {H.}~\bibnamefont {Polshyn}}, \bibinfo
  {author} {\bibfnamefont {Y.}~\bibnamefont {Zhang}}, \bibinfo {author}
  {\bibfnamefont {J.}~\bibnamefont {Zhu}}, \bibinfo {author} {\bibfnamefont
  {K.}~\bibnamefont {Watanabe}}, \bibinfo {author} {\bibfnamefont
  {T.}~\bibnamefont {Taniguchi}}, \bibinfo {author} {\bibfnamefont
  {L.}~\bibnamefont {Balents}},\ and\ \bibinfo {author} {\bibfnamefont
  {A.}~\bibnamefont {Young}},\ }\bibfield  {title} {\bibinfo {title} {Intrinsic
  quantized anomalous hall effect in a moir{\'e} heterostructure},\ }\href@noop
  {} {\bibfield  {journal} {\bibinfo  {journal} {Science}\ }\textbf {\bibinfo
  {volume} {367}},\ \bibinfo {pages} {900} (\bibinfo {year}
  {2020})}\BibitemShut {NoStop}%
\bibitem [{\citenamefont {Li}\ \emph {et~al.}(2021{\natexlab{b}})\citenamefont
  {Li}, \citenamefont {Jiang}, \citenamefont {Shen}, \citenamefont {Zhang},
  \citenamefont {Li}, \citenamefont {Tao}, \citenamefont {Devakul},
  \citenamefont {Watanabe}, \citenamefont {Taniguchi}, \citenamefont {Fu} \emph
  {et~al.}}]{li2021quantum}%
  \BibitemOpen
  \bibfield  {author} {\bibinfo {author} {\bibfnamefont {T.}~\bibnamefont
  {Li}}, \bibinfo {author} {\bibfnamefont {S.}~\bibnamefont {Jiang}}, \bibinfo
  {author} {\bibfnamefont {B.}~\bibnamefont {Shen}}, \bibinfo {author}
  {\bibfnamefont {Y.}~\bibnamefont {Zhang}}, \bibinfo {author} {\bibfnamefont
  {L.}~\bibnamefont {Li}}, \bibinfo {author} {\bibfnamefont {Z.}~\bibnamefont
  {Tao}}, \bibinfo {author} {\bibfnamefont {T.}~\bibnamefont {Devakul}},
  \bibinfo {author} {\bibfnamefont {K.}~\bibnamefont {Watanabe}}, \bibinfo
  {author} {\bibfnamefont {T.}~\bibnamefont {Taniguchi}}, \bibinfo {author}
  {\bibfnamefont {L.}~\bibnamefont {Fu}}, \emph {et~al.},\ }\bibfield  {title}
  {\bibinfo {title} {Quantum anomalous hall effect from intertwined moir{\'e}
  bands},\ }\href@noop {} {\bibfield  {journal} {\bibinfo  {journal} {Nature}\
  }\textbf {\bibinfo {volume} {600}},\ \bibinfo {pages} {641} (\bibinfo {year}
  {2021}{\natexlab{b}})}\BibitemShut {NoStop}%
\bibitem [{\citenamefont {Lopes~dos Santos}\ \emph {et~al.}(2007)\citenamefont
  {Lopes~dos Santos}, \citenamefont {Peres},\ and\ \citenamefont
  {Castro~Neto}}]{PhysRevLett.99.256802}%
  \BibitemOpen
  \bibfield  {author} {\bibinfo {author} {\bibfnamefont {J.~M.~B.}\
  \bibnamefont {Lopes~dos Santos}}, \bibinfo {author} {\bibfnamefont
  {N.~M.~R.}\ \bibnamefont {Peres}},\ and\ \bibinfo {author} {\bibfnamefont
  {A.~H.}\ \bibnamefont {Castro~Neto}},\ }\bibfield  {title} {\bibinfo {title}
  {Graphene bilayer with a twist: Electronic structure},\ }\href
  {https://doi.org/10.1103/PhysRevLett.99.256802} {\bibfield  {journal}
  {\bibinfo  {journal} {Phys. Rev. Lett.}\ }\textbf {\bibinfo {volume} {99}},\
  \bibinfo {pages} {256802} (\bibinfo {year} {2007})}\BibitemShut {NoStop}%
\bibitem [{\citenamefont {Trambly~de Laissardi{\`e}re}\ \emph
  {et~al.}(2010)\citenamefont {Trambly~de Laissardi{\`e}re}, \citenamefont
  {Mayou},\ and\ \citenamefont {Magaud}}]{TramblyMagaud2010}%
  \BibitemOpen
  \bibfield  {author} {\bibinfo {author} {\bibfnamefont {G.}~\bibnamefont
  {Trambly~de Laissardi{\`e}re}}, \bibinfo {author} {\bibfnamefont
  {D.}~\bibnamefont {Mayou}},\ and\ \bibinfo {author} {\bibfnamefont
  {L.}~\bibnamefont {Magaud}},\ }\bibfield  {title} {\bibinfo {title}
  {Localization of dirac electrons in rotated graphene bilayers},\ }\href@noop
  {} {\bibfield  {journal} {\bibinfo  {journal} {Nano letters}\ }\textbf
  {\bibinfo {volume} {10}},\ \bibinfo {pages} {804} (\bibinfo {year}
  {2010})}\BibitemShut {NoStop}%
\bibitem [{\citenamefont {Su\'arez~Morell}\ \emph {et~al.}(2010)\citenamefont
  {Su\'arez~Morell}, \citenamefont {Correa}, \citenamefont {Vargas},
  \citenamefont {Pacheco},\ and\ \citenamefont {Barticevic}}]{Morell2011}%
  \BibitemOpen
  \bibfield  {author} {\bibinfo {author} {\bibfnamefont {E.}~\bibnamefont
  {Su\'arez~Morell}}, \bibinfo {author} {\bibfnamefont {J.~D.}\ \bibnamefont
  {Correa}}, \bibinfo {author} {\bibfnamefont {P.}~\bibnamefont {Vargas}},
  \bibinfo {author} {\bibfnamefont {M.}~\bibnamefont {Pacheco}},\ and\ \bibinfo
  {author} {\bibfnamefont {Z.}~\bibnamefont {Barticevic}},\ }\bibfield  {title}
  {\bibinfo {title} {Flat bands in slightly twisted bilayer graphene:
  Tight-binding calculations},\ }\href
  {https://doi.org/10.1103/PhysRevB.82.121407} {\bibfield  {journal} {\bibinfo
  {journal} {Phys. Rev. B}\ }\textbf {\bibinfo {volume} {82}},\ \bibinfo
  {pages} {121407} (\bibinfo {year} {2010})}\BibitemShut {NoStop}%
\bibitem [{\citenamefont {Bistritzer}\ and\ \citenamefont
  {MacDonald}(2011)}]{BistritzerMacDonald2011}%
  \BibitemOpen
  \bibfield  {author} {\bibinfo {author} {\bibfnamefont {R.}~\bibnamefont
  {Bistritzer}}\ and\ \bibinfo {author} {\bibfnamefont {A.~H.}\ \bibnamefont
  {MacDonald}},\ }\bibfield  {title} {\bibinfo {title} {Moir{\'e} bands in
  twisted double-layer graphene},\ }\href@noop {} {\bibfield  {journal}
  {\bibinfo  {journal} {Proceedings of the National Academy of Sciences}\
  }\textbf {\bibinfo {volume} {108}},\ \bibinfo {pages} {12233} (\bibinfo
  {year} {2011})}\BibitemShut {NoStop}%
\bibitem [{\citenamefont {Wu}\ \emph {et~al.}(2018)\citenamefont {Wu},
  \citenamefont {Lovorn}, \citenamefont {Tutuc},\ and\ \citenamefont
  {MacDonald}}]{PhysRevLett.121.026402}%
  \BibitemOpen
  \bibfield  {author} {\bibinfo {author} {\bibfnamefont {F.}~\bibnamefont
  {Wu}}, \bibinfo {author} {\bibfnamefont {T.}~\bibnamefont {Lovorn}}, \bibinfo
  {author} {\bibfnamefont {E.}~\bibnamefont {Tutuc}},\ and\ \bibinfo {author}
  {\bibfnamefont {A.~H.}\ \bibnamefont {MacDonald}},\ }\bibfield  {title}
  {\bibinfo {title} {Hubbard model physics in transition metal dichalcogenide
  moir\'e bands},\ }\href {https://doi.org/10.1103/PhysRevLett.121.026402}
  {\bibfield  {journal} {\bibinfo  {journal} {Phys. Rev. Lett.}\ }\textbf
  {\bibinfo {volume} {121}},\ \bibinfo {pages} {026402} (\bibinfo {year}
  {2018})}\BibitemShut {NoStop}%
\bibitem [{\citenamefont {Tarnopolsky}\ \emph {et~al.}(2019)\citenamefont
  {Tarnopolsky}, \citenamefont {Kruchkov},\ and\ \citenamefont
  {Vishwanath}}]{PhysRevLett.122.106405}%
  \BibitemOpen
  \bibfield  {author} {\bibinfo {author} {\bibfnamefont {G.}~\bibnamefont
  {Tarnopolsky}}, \bibinfo {author} {\bibfnamefont {A.~J.}\ \bibnamefont
  {Kruchkov}},\ and\ \bibinfo {author} {\bibfnamefont {A.}~\bibnamefont
  {Vishwanath}},\ }\bibfield  {title} {\bibinfo {title} {Origin of magic angles
  in twisted bilayer graphene},\ }\href
  {https://doi.org/10.1103/PhysRevLett.122.106405} {\bibfield  {journal}
  {\bibinfo  {journal} {Phys. Rev. Lett.}\ }\textbf {\bibinfo {volume} {122}},\
  \bibinfo {pages} {106405} (\bibinfo {year} {2019})}\BibitemShut {NoStop}%
\bibitem [{\citenamefont {Lopes~dos Santos}\ \emph {et~al.}(2012)\citenamefont
  {Lopes~dos Santos}, \citenamefont {Peres},\ and\ \citenamefont
  {Castro~Neto}}]{PhysRevB.86.155449}%
  \BibitemOpen
  \bibfield  {author} {\bibinfo {author} {\bibfnamefont {J.~M.~B.}\
  \bibnamefont {Lopes~dos Santos}}, \bibinfo {author} {\bibfnamefont
  {N.~M.~R.}\ \bibnamefont {Peres}},\ and\ \bibinfo {author} {\bibfnamefont
  {A.~H.}\ \bibnamefont {Castro~Neto}},\ }\bibfield  {title} {\bibinfo {title}
  {Continuum model of the twisted graphene bilayer},\ }\href
  {https://doi.org/10.1103/PhysRevB.86.155449} {\bibfield  {journal} {\bibinfo
  {journal} {Phys. Rev. B}\ }\textbf {\bibinfo {volume} {86}},\ \bibinfo
  {pages} {155449} (\bibinfo {year} {2012})}\BibitemShut {NoStop}%
\bibitem [{\citenamefont {Vafek}\ and\ \citenamefont
  {Kang}(2020)}]{PhysRevLett.125.257602}%
  \BibitemOpen
  \bibfield  {author} {\bibinfo {author} {\bibfnamefont {O.}~\bibnamefont
  {Vafek}}\ and\ \bibinfo {author} {\bibfnamefont {J.}~\bibnamefont {Kang}},\
  }\bibfield  {title} {\bibinfo {title} {Renormalization group study of hidden
  symmetry in twisted bilayer graphene with coulomb interactions},\ }\href
  {https://doi.org/10.1103/PhysRevLett.125.257602} {\bibfield  {journal}
  {\bibinfo  {journal} {Phys. Rev. Lett.}\ }\textbf {\bibinfo {volume} {125}},\
  \bibinfo {pages} {257602} (\bibinfo {year} {2020})}\BibitemShut {NoStop}%
\bibitem [{\citenamefont {Bernevig}\ \emph
  {et~al.}(2021{\natexlab{a}})\citenamefont {Bernevig}, \citenamefont {Song},
  \citenamefont {Regnault},\ and\ \citenamefont {Lian}}]{PhysRevB.103.205411}%
  \BibitemOpen
  \bibfield  {author} {\bibinfo {author} {\bibfnamefont {B.~A.}\ \bibnamefont
  {Bernevig}}, \bibinfo {author} {\bibfnamefont {Z.-D.}\ \bibnamefont {Song}},
  \bibinfo {author} {\bibfnamefont {N.}~\bibnamefont {Regnault}},\ and\
  \bibinfo {author} {\bibfnamefont {B.}~\bibnamefont {Lian}},\ }\bibfield
  {title} {\bibinfo {title} {Twisted bilayer graphene. i. matrix elements,
  approximations, perturbation theory, and a
  $k\ifmmode\cdot\else\textperiodcentered\fi{}p$ two-band model},\ }\href
  {https://doi.org/10.1103/PhysRevB.103.205411} {\bibfield  {journal} {\bibinfo
   {journal} {Phys. Rev. B}\ }\textbf {\bibinfo {volume} {103}},\ \bibinfo
  {pages} {205411} (\bibinfo {year} {2021}{\natexlab{a}})}\BibitemShut
  {NoStop}%
\bibitem [{\citenamefont {Song}\ \emph {et~al.}(2021)\citenamefont {Song},
  \citenamefont {Lian}, \citenamefont {Regnault},\ and\ \citenamefont
  {Bernevig}}]{PhysRevB.103.205412}%
  \BibitemOpen
  \bibfield  {author} {\bibinfo {author} {\bibfnamefont {Z.-D.}\ \bibnamefont
  {Song}}, \bibinfo {author} {\bibfnamefont {B.}~\bibnamefont {Lian}}, \bibinfo
  {author} {\bibfnamefont {N.}~\bibnamefont {Regnault}},\ and\ \bibinfo
  {author} {\bibfnamefont {B.~A.}\ \bibnamefont {Bernevig}},\ }\bibfield
  {title} {\bibinfo {title} {Twisted bilayer graphene. ii. stable symmetry
  anomaly},\ }\href {https://doi.org/10.1103/PhysRevB.103.205412} {\bibfield
  {journal} {\bibinfo  {journal} {Phys. Rev. B}\ }\textbf {\bibinfo {volume}
  {103}},\ \bibinfo {pages} {205412} (\bibinfo {year} {2021})}\BibitemShut
  {NoStop}%
\bibitem [{\citenamefont {Bernevig}\ \emph
  {et~al.}(2021{\natexlab{b}})\citenamefont {Bernevig}, \citenamefont {Song},
  \citenamefont {Regnault},\ and\ \citenamefont {Lian}}]{PhysRevB.103.205413}%
  \BibitemOpen
  \bibfield  {author} {\bibinfo {author} {\bibfnamefont {B.~A.}\ \bibnamefont
  {Bernevig}}, \bibinfo {author} {\bibfnamefont {Z.-D.}\ \bibnamefont {Song}},
  \bibinfo {author} {\bibfnamefont {N.}~\bibnamefont {Regnault}},\ and\
  \bibinfo {author} {\bibfnamefont {B.}~\bibnamefont {Lian}},\ }\bibfield
  {title} {\bibinfo {title} {Twisted bilayer graphene. iii. interacting
  hamiltonian and exact symmetries},\ }\href
  {https://doi.org/10.1103/PhysRevB.103.205413} {\bibfield  {journal} {\bibinfo
   {journal} {Phys. Rev. B}\ }\textbf {\bibinfo {volume} {103}},\ \bibinfo
  {pages} {205413} (\bibinfo {year} {2021}{\natexlab{b}})}\BibitemShut
  {NoStop}%
\bibitem [{\citenamefont {Lian}\ \emph {et~al.}(2021)\citenamefont {Lian},
  \citenamefont {Song}, \citenamefont {Regnault}, \citenamefont {Efetov},
  \citenamefont {Yazdani},\ and\ \citenamefont
  {Bernevig}}]{PhysRevB.103.205414}%
  \BibitemOpen
  \bibfield  {author} {\bibinfo {author} {\bibfnamefont {B.}~\bibnamefont
  {Lian}}, \bibinfo {author} {\bibfnamefont {Z.-D.}\ \bibnamefont {Song}},
  \bibinfo {author} {\bibfnamefont {N.}~\bibnamefont {Regnault}}, \bibinfo
  {author} {\bibfnamefont {D.~K.}\ \bibnamefont {Efetov}}, \bibinfo {author}
  {\bibfnamefont {A.}~\bibnamefont {Yazdani}},\ and\ \bibinfo {author}
  {\bibfnamefont {B.~A.}\ \bibnamefont {Bernevig}},\ }\bibfield  {title}
  {\bibinfo {title} {Twisted bilayer graphene. iv. exact insulator ground
  states and phase diagram},\ }\href
  {https://doi.org/10.1103/PhysRevB.103.205414} {\bibfield  {journal} {\bibinfo
   {journal} {Phys. Rev. B}\ }\textbf {\bibinfo {volume} {103}},\ \bibinfo
  {pages} {205414} (\bibinfo {year} {2021})}\BibitemShut {NoStop}%
\bibitem [{\citenamefont {Bernevig}\ \emph
  {et~al.}(2021{\natexlab{c}})\citenamefont {Bernevig}, \citenamefont {Lian},
  \citenamefont {Cowsik}, \citenamefont {Xie}, \citenamefont {Regnault},\ and\
  \citenamefont {Song}}]{PhysRevB.103.205415}%
  \BibitemOpen
  \bibfield  {author} {\bibinfo {author} {\bibfnamefont {B.~A.}\ \bibnamefont
  {Bernevig}}, \bibinfo {author} {\bibfnamefont {B.}~\bibnamefont {Lian}},
  \bibinfo {author} {\bibfnamefont {A.}~\bibnamefont {Cowsik}}, \bibinfo
  {author} {\bibfnamefont {F.}~\bibnamefont {Xie}}, \bibinfo {author}
  {\bibfnamefont {N.}~\bibnamefont {Regnault}},\ and\ \bibinfo {author}
  {\bibfnamefont {Z.-D.}\ \bibnamefont {Song}},\ }\bibfield  {title} {\bibinfo
  {title} {Twisted bilayer graphene. v. exact analytic many-body excitations in
  coulomb hamiltonians: Charge gap, goldstone modes, and absence of cooper
  pairing},\ }\href {https://doi.org/10.1103/PhysRevB.103.205415} {\bibfield
  {journal} {\bibinfo  {journal} {Phys. Rev. B}\ }\textbf {\bibinfo {volume}
  {103}},\ \bibinfo {pages} {205415} (\bibinfo {year}
  {2021}{\natexlab{c}})}\BibitemShut {NoStop}%
\bibitem [{\citenamefont {Xie}\ \emph {et~al.}(2021)\citenamefont {Xie},
  \citenamefont {Cowsik}, \citenamefont {Song}, \citenamefont {Lian},
  \citenamefont {Bernevig},\ and\ \citenamefont
  {Regnault}}]{PhysRevB.103.205416}%
  \BibitemOpen
  \bibfield  {author} {\bibinfo {author} {\bibfnamefont {F.}~\bibnamefont
  {Xie}}, \bibinfo {author} {\bibfnamefont {A.}~\bibnamefont {Cowsik}},
  \bibinfo {author} {\bibfnamefont {Z.-D.}\ \bibnamefont {Song}}, \bibinfo
  {author} {\bibfnamefont {B.}~\bibnamefont {Lian}}, \bibinfo {author}
  {\bibfnamefont {B.~A.}\ \bibnamefont {Bernevig}},\ and\ \bibinfo {author}
  {\bibfnamefont {N.}~\bibnamefont {Regnault}},\ }\bibfield  {title} {\bibinfo
  {title} {Twisted bilayer graphene. vi. an exact diagonalization study at
  nonzero integer filling},\ }\href
  {https://doi.org/10.1103/PhysRevB.103.205416} {\bibfield  {journal} {\bibinfo
   {journal} {Phys. Rev. B}\ }\textbf {\bibinfo {volume} {103}},\ \bibinfo
  {pages} {205416} (\bibinfo {year} {2021})}\BibitemShut {NoStop}%
\bibitem [{\citenamefont {Pixley}\ \emph {et~al.}(2018)\citenamefont {Pixley},
  \citenamefont {Wilson}, \citenamefont {Huse},\ and\ \citenamefont
  {Gopalakrishnan}}]{PixleyGopalakrishnan2018}%
  \BibitemOpen
  \bibfield  {author} {\bibinfo {author} {\bibfnamefont {J.~H.}\ \bibnamefont
  {Pixley}}, \bibinfo {author} {\bibfnamefont {J.~H.}\ \bibnamefont {Wilson}},
  \bibinfo {author} {\bibfnamefont {D.~A.}\ \bibnamefont {Huse}},\ and\
  \bibinfo {author} {\bibfnamefont {S.}~\bibnamefont {Gopalakrishnan}},\
  }\bibfield  {title} {\bibinfo {title} {Weyl semimetal to metal phase
  transitions driven by quasiperiodic potentials},\ }\href
  {https://doi.org/10.1103/PhysRevLett.120.207604} {\bibfield  {journal}
  {\bibinfo  {journal} {Phys. Rev. Lett.}\ }\textbf {\bibinfo {volume} {120}},\
  \bibinfo {pages} {207604} (\bibinfo {year} {2018})}\BibitemShut {NoStop}%
\bibitem [{\citenamefont {Fu}\ \emph {et~al.}(2020)\citenamefont {Fu},
  \citenamefont {K{\"o}nig}, \citenamefont {Wilson}, \citenamefont {Chou},\
  and\ \citenamefont {Pixley}}]{FuPixley2020}%
  \BibitemOpen
  \bibfield  {author} {\bibinfo {author} {\bibfnamefont {Y.}~\bibnamefont
  {Fu}}, \bibinfo {author} {\bibfnamefont {E.~J.}\ \bibnamefont {K{\"o}nig}},
  \bibinfo {author} {\bibfnamefont {J.~H.}\ \bibnamefont {Wilson}}, \bibinfo
  {author} {\bibfnamefont {Y.-Z.}\ \bibnamefont {Chou}},\ and\ \bibinfo
  {author} {\bibfnamefont {J.~H.}\ \bibnamefont {Pixley}},\ }\bibfield  {title}
  {\bibinfo {title} {Magic-angle semimetals},\ }\href@noop {} {\bibfield
  {journal} {\bibinfo  {journal} {npj Quantum Materials}\ }\textbf {\bibinfo
  {volume} {5}},\ \bibinfo {pages} {1} (\bibinfo {year} {2020})}\BibitemShut
  {NoStop}%
\bibitem [{\citenamefont {Chou}\ \emph {et~al.}(2020)\citenamefont {Chou},
  \citenamefont {Fu}, \citenamefont {Wilson}, \citenamefont {K\"onig},\ and\
  \citenamefont {Pixley}}]{ChouPixley2020}%
  \BibitemOpen
  \bibfield  {author} {\bibinfo {author} {\bibfnamefont {Y.-Z.}\ \bibnamefont
  {Chou}}, \bibinfo {author} {\bibfnamefont {Y.}~\bibnamefont {Fu}}, \bibinfo
  {author} {\bibfnamefont {J.~H.}\ \bibnamefont {Wilson}}, \bibinfo {author}
  {\bibfnamefont {E.~J.}\ \bibnamefont {K\"onig}},\ and\ \bibinfo {author}
  {\bibfnamefont {J.~H.}\ \bibnamefont {Pixley}},\ }\bibfield  {title}
  {\bibinfo {title} {Magic-angle semimetals with chiral symmetry},\ }\href
  {https://doi.org/10.1103/PhysRevB.101.235121} {\bibfield  {journal} {\bibinfo
   {journal} {Phys. Rev. B}\ }\textbf {\bibinfo {volume} {101}},\ \bibinfo
  {pages} {235121} (\bibinfo {year} {2020})}\BibitemShut {NoStop}%
\bibitem [{\citenamefont {Luo}\ and\ \citenamefont
  {Zhang}(2021)}]{PhysRevLett.126.103201}%
  \BibitemOpen
  \bibfield  {author} {\bibinfo {author} {\bibfnamefont {X.-W.}\ \bibnamefont
  {Luo}}\ and\ \bibinfo {author} {\bibfnamefont {C.}~\bibnamefont {Zhang}},\
  }\bibfield  {title} {\bibinfo {title} {Spin-twisted optical lattices: Tunable
  flat bands and larkin-ovchinnikov superfluids},\ }\href
  {https://doi.org/10.1103/PhysRevLett.126.103201} {\bibfield  {journal}
  {\bibinfo  {journal} {Phys. Rev. Lett.}\ }\textbf {\bibinfo {volume} {126}},\
  \bibinfo {pages} {103201} (\bibinfo {year} {2021})}\BibitemShut {NoStop}%
\bibitem [{\citenamefont {Gonz\'alez-Tudela}\ and\ \citenamefont
  {Cirac}(2019)}]{PhysRevA.100.053604}%
  \BibitemOpen
  \bibfield  {author} {\bibinfo {author} {\bibfnamefont {A.}~\bibnamefont
  {Gonz\'alez-Tudela}}\ and\ \bibinfo {author} {\bibfnamefont {J.~I.}\
  \bibnamefont {Cirac}},\ }\bibfield  {title} {\bibinfo {title} {Cold atoms in
  twisted-bilayer optical potentials},\ }\href
  {https://doi.org/10.1103/PhysRevA.100.053604} {\bibfield  {journal} {\bibinfo
   {journal} {Phys. Rev. A}\ }\textbf {\bibinfo {volume} {100}},\ \bibinfo
  {pages} {053604} (\bibinfo {year} {2019})}\BibitemShut {NoStop}%
\bibitem [{\citenamefont {Salamon}\ \emph
  {et~al.}(2020{\natexlab{a}})\citenamefont {Salamon}, \citenamefont {Celi},
  \citenamefont {Chhajlany}, \citenamefont {Frerot}, \citenamefont
  {Lewenstein}, \citenamefont {Tarruell},\ and\ \citenamefont
  {Rakshit}}]{PhysRevLett.125.030504}%
  \BibitemOpen
  \bibfield  {author} {\bibinfo {author} {\bibfnamefont {T.}~\bibnamefont
  {Salamon}}, \bibinfo {author} {\bibfnamefont {A.}~\bibnamefont {Celi}},
  \bibinfo {author} {\bibfnamefont {R.~W.}\ \bibnamefont {Chhajlany}}, \bibinfo
  {author} {\bibfnamefont {I.}~\bibnamefont {Frerot}}, \bibinfo {author}
  {\bibfnamefont {M.}~\bibnamefont {Lewenstein}}, \bibinfo {author}
  {\bibfnamefont {L.}~\bibnamefont {Tarruell}},\ and\ \bibinfo {author}
  {\bibfnamefont {D.}~\bibnamefont {Rakshit}},\ }\bibfield  {title} {\bibinfo
  {title} {Simulating twistronics without a twist},\ }\href
  {https://doi.org/10.1103/PhysRevLett.125.030504} {\bibfield  {journal}
  {\bibinfo  {journal} {Phys. Rev. Lett.}\ }\textbf {\bibinfo {volume} {125}},\
  \bibinfo {pages} {030504} (\bibinfo {year} {2020}{\natexlab{a}})}\BibitemShut
  {NoStop}%
\bibitem [{\citenamefont {Salamon}\ \emph
  {et~al.}(2020{\natexlab{b}})\citenamefont {Salamon}, \citenamefont
  {Chhajlany}, \citenamefont {Dauphin}, \citenamefont {Lewenstein},\ and\
  \citenamefont {Rakshit}}]{PhysRevB.102.235126}%
  \BibitemOpen
  \bibfield  {author} {\bibinfo {author} {\bibfnamefont {T.}~\bibnamefont
  {Salamon}}, \bibinfo {author} {\bibfnamefont {R.~W.}\ \bibnamefont
  {Chhajlany}}, \bibinfo {author} {\bibfnamefont {A.}~\bibnamefont {Dauphin}},
  \bibinfo {author} {\bibfnamefont {M.}~\bibnamefont {Lewenstein}},\ and\
  \bibinfo {author} {\bibfnamefont {D.}~\bibnamefont {Rakshit}},\ }\bibfield
  {title} {\bibinfo {title} {Quantum anomalous hall phase in synthetic bilayers
  via twistronics without a twist},\ }\href
  {https://doi.org/10.1103/PhysRevB.102.235126} {\bibfield  {journal} {\bibinfo
   {journal} {Phys. Rev. B}\ }\textbf {\bibinfo {volume} {102}},\ \bibinfo
  {pages} {235126} (\bibinfo {year} {2020}{\natexlab{b}})}\BibitemShut
  {NoStop}%
\bibitem [{\citenamefont {Fu}\ \emph {et~al.}(2021)\citenamefont {Fu},
  \citenamefont {Wilson},\ and\ \citenamefont {Pixley}}]{PhysRevB.104.L041106}%
  \BibitemOpen
  \bibfield  {author} {\bibinfo {author} {\bibfnamefont {Y.}~\bibnamefont
  {Fu}}, \bibinfo {author} {\bibfnamefont {J.~H.}\ \bibnamefont {Wilson}},\
  and\ \bibinfo {author} {\bibfnamefont {J.~H.}\ \bibnamefont {Pixley}},\
  }\bibfield  {title} {\bibinfo {title} {Flat topological bands and eigenstate
  criticality in a quasiperiodic insulator},\ }\href
  {https://doi.org/10.1103/PhysRevB.104.L041106} {\bibfield  {journal}
  {\bibinfo  {journal} {Phys. Rev. B}\ }\textbf {\bibinfo {volume} {104}},\
  \bibinfo {pages} {L041106} (\bibinfo {year} {2021})}\BibitemShut {NoStop}%
\bibitem [{\citenamefont {Cano}\ \emph {et~al.}(2021)\citenamefont {Cano},
  \citenamefont {Fang}, \citenamefont {Pixley},\ and\ \citenamefont
  {Wilson}}]{PhysRevB.103.155157}%
  \BibitemOpen
  \bibfield  {author} {\bibinfo {author} {\bibfnamefont {J.}~\bibnamefont
  {Cano}}, \bibinfo {author} {\bibfnamefont {S.}~\bibnamefont {Fang}}, \bibinfo
  {author} {\bibfnamefont {J.~H.}\ \bibnamefont {Pixley}},\ and\ \bibinfo
  {author} {\bibfnamefont {J.~H.}\ \bibnamefont {Wilson}},\ }\bibfield  {title}
  {\bibinfo {title} {Moir\'e superlattice on the surface of a topological
  insulator},\ }\href {https://doi.org/10.1103/PhysRevB.103.155157} {\bibfield
  {journal} {\bibinfo  {journal} {Phys. Rev. B}\ }\textbf {\bibinfo {volume}
  {103}},\ \bibinfo {pages} {155157} (\bibinfo {year} {2021})}\BibitemShut
  {NoStop}%
\bibitem [{\citenamefont {Wang}\ \emph
  {et~al.}(2021{\natexlab{a}})\citenamefont {Wang}, \citenamefont {Yuan},\ and\
  \citenamefont {Fu}}]{PhysRevX.11.021024}%
  \BibitemOpen
  \bibfield  {author} {\bibinfo {author} {\bibfnamefont {T.}~\bibnamefont
  {Wang}}, \bibinfo {author} {\bibfnamefont {N.~F.~Q.}\ \bibnamefont {Yuan}},\
  and\ \bibinfo {author} {\bibfnamefont {L.}~\bibnamefont {Fu}},\ }\bibfield
  {title} {\bibinfo {title} {Moir\'e surface states and enhanced
  superconductivity in topological insulators},\ }\href
  {https://doi.org/10.1103/PhysRevX.11.021024} {\bibfield  {journal} {\bibinfo
  {journal} {Phys. Rev. X}\ }\textbf {\bibinfo {volume} {11}},\ \bibinfo
  {pages} {021024} (\bibinfo {year} {2021}{\natexlab{a}})}\BibitemShut
  {NoStop}%
\bibitem [{\citenamefont {Chou}\ \emph {et~al.}(2021)\citenamefont {Chou},
  \citenamefont {Cano},\ and\ \citenamefont {Pixley}}]{PhysRevB.104.L201113}%
  \BibitemOpen
  \bibfield  {author} {\bibinfo {author} {\bibfnamefont {Y.-Z.}\ \bibnamefont
  {Chou}}, \bibinfo {author} {\bibfnamefont {J.}~\bibnamefont {Cano}},\ and\
  \bibinfo {author} {\bibfnamefont {J.~H.}\ \bibnamefont {Pixley}},\ }\bibfield
   {title} {\bibinfo {title} {Band manipulation and spin texture in interacting
  moir\'e helical edges},\ }\href
  {https://doi.org/10.1103/PhysRevB.104.L201113} {\bibfield  {journal}
  {\bibinfo  {journal} {Phys. Rev. B}\ }\textbf {\bibinfo {volume} {104}},\
  \bibinfo {pages} {L201113} (\bibinfo {year} {2021})}\BibitemShut {NoStop}%
\bibitem [{\citenamefont {Volkov}\ \emph {et~al.}(2020)\citenamefont {Volkov},
  \citenamefont {Wilson},\ and\ \citenamefont {Pixley}}]{volkov2020magic}%
  \BibitemOpen
  \bibfield  {author} {\bibinfo {author} {\bibfnamefont {P.~A.}\ \bibnamefont
  {Volkov}}, \bibinfo {author} {\bibfnamefont {J.~H.}\ \bibnamefont {Wilson}},\
  and\ \bibinfo {author} {\bibfnamefont {J.}~\bibnamefont {Pixley}},\
  }\bibfield  {title} {\bibinfo {title} {Magic angles and current-induced
  topology in twisted nodal superconductors},\ }\href@noop {} {\bibfield
  {journal} {\bibinfo  {journal} {arXiv preprint arXiv:2012.07860}\ } (\bibinfo
  {year} {2020})}\BibitemShut {NoStop}%
\bibitem [{\citenamefont {Tummuru}\ \emph {et~al.}(2022)\citenamefont
  {Tummuru}, \citenamefont {Plugge},\ and\ \citenamefont
  {Franz}}]{tummuru2022josephson}%
  \BibitemOpen
  \bibfield  {author} {\bibinfo {author} {\bibfnamefont {T.}~\bibnamefont
  {Tummuru}}, \bibinfo {author} {\bibfnamefont {S.}~\bibnamefont {Plugge}},\
  and\ \bibinfo {author} {\bibfnamefont {M.}~\bibnamefont {Franz}},\ }\bibfield
   {title} {\bibinfo {title} {Josephson effects in twisted cuprate bilayers},\
  }\href@noop {} {\bibfield  {journal} {\bibinfo  {journal} {Physical Review
  B}\ }\textbf {\bibinfo {volume} {105}},\ \bibinfo {pages} {064501} (\bibinfo
  {year} {2022})}\BibitemShut {NoStop}%
\bibitem [{\citenamefont {Lee}\ \emph {et~al.}(2019)\citenamefont {Lee},
  \citenamefont {Khalaf}, \citenamefont {Liu}, \citenamefont {Liu},
  \citenamefont {Hao}, \citenamefont {Kim},\ and\ \citenamefont
  {Vishwanath}}]{lee2019theory}%
  \BibitemOpen
  \bibfield  {author} {\bibinfo {author} {\bibfnamefont {J.~Y.}\ \bibnamefont
  {Lee}}, \bibinfo {author} {\bibfnamefont {E.}~\bibnamefont {Khalaf}},
  \bibinfo {author} {\bibfnamefont {S.}~\bibnamefont {Liu}}, \bibinfo {author}
  {\bibfnamefont {X.}~\bibnamefont {Liu}}, \bibinfo {author} {\bibfnamefont
  {Z.}~\bibnamefont {Hao}}, \bibinfo {author} {\bibfnamefont {P.}~\bibnamefont
  {Kim}},\ and\ \bibinfo {author} {\bibfnamefont {A.}~\bibnamefont
  {Vishwanath}},\ }\bibfield  {title} {\bibinfo {title} {Theory of correlated
  insulating behaviour and spin-triplet superconductivity in twisted double
  bilayer graphene},\ }\href@noop {} {\bibfield  {journal} {\bibinfo  {journal}
  {Nature communications}\ }\textbf {\bibinfo {volume} {10}},\ \bibinfo {pages}
  {1} (\bibinfo {year} {2019})}\BibitemShut {NoStop}%
\bibitem [{\citenamefont {Lee}\ and\ \citenamefont
  {Pixley}(2021)}]{LeePixley2021}%
  \BibitemOpen
  \bibfield  {author} {\bibinfo {author} {\bibfnamefont {J.}~\bibnamefont
  {Lee}}\ and\ \bibinfo {author} {\bibfnamefont {J.}~\bibnamefont {Pixley}},\
  }\bibfield  {title} {\bibinfo {title} {Emulating twisted double bilayer
  graphene with a multiorbital optical lattice},\ }\href@noop {} {\bibfield
  {journal} {\bibinfo  {journal} {arXiv preprint arXiv:2112.13797}\ } (\bibinfo
  {year} {2021})}\BibitemShut {NoStop}%
\bibitem [{\citenamefont {Ledwith}\ \emph {et~al.}(2021)\citenamefont
  {Ledwith}, \citenamefont {Khalaf}, \citenamefont {Zhu}, \citenamefont {Carr},
  \citenamefont {Kaxiras},\ and\ \citenamefont {Vishwanath}}]{ledwith2021tb}%
  \BibitemOpen
  \bibfield  {author} {\bibinfo {author} {\bibfnamefont {P.~J.}\ \bibnamefont
  {Ledwith}}, \bibinfo {author} {\bibfnamefont {E.}~\bibnamefont {Khalaf}},
  \bibinfo {author} {\bibfnamefont {Z.}~\bibnamefont {Zhu}}, \bibinfo {author}
  {\bibfnamefont {S.}~\bibnamefont {Carr}}, \bibinfo {author} {\bibfnamefont
  {E.}~\bibnamefont {Kaxiras}},\ and\ \bibinfo {author} {\bibfnamefont
  {A.}~\bibnamefont {Vishwanath}},\ }\bibfield  {title} {\bibinfo {title} {Tb
  or not tb? contrasting properties of twisted bilayer graphene and the
  alternating twist $ n $-layer structures ($ n= 3, 4, 5,\dots$)},\ }\href@noop
  {} {\bibfield  {journal} {\bibinfo  {journal} {arXiv preprint
  arXiv:2111.11060}\ } (\bibinfo {year} {2021})}\BibitemShut {NoStop}%
\bibitem [{\citenamefont {Gonzalez-Tudela}\ and\ \citenamefont
  {Cirac}(2019)}]{GonzalezCirac2019}%
  \BibitemOpen
  \bibfield  {author} {\bibinfo {author} {\bibfnamefont {A.}~\bibnamefont
  {Gonzalez-Tudela}}\ and\ \bibinfo {author} {\bibfnamefont {J.~I.}\
  \bibnamefont {Cirac}},\ }\bibfield  {title} {\bibinfo {title} {Cold atoms in
  twisted-bilayer optical potentials},\ }\href
  {https://doi.org/10.1103/PhysRevA.100.053604} {\bibfield  {journal} {\bibinfo
   {journal} {Phys. Rev. A}\ }\textbf {\bibinfo {volume} {100}},\ \bibinfo
  {pages} {053604} (\bibinfo {year} {2019})}\BibitemShut {NoStop}%
\bibitem [{\citenamefont {Salamon}\ \emph
  {et~al.}(2020{\natexlab{c}})\citenamefont {Salamon}, \citenamefont {Celi},
  \citenamefont {Chhajlany}, \citenamefont {Fr\'erot}, \citenamefont
  {Lewenstein}, \citenamefont {Tarruell},\ and\ \citenamefont
  {Rakshit}}]{SalamonRakshit2020}%
  \BibitemOpen
  \bibfield  {author} {\bibinfo {author} {\bibfnamefont {T.}~\bibnamefont
  {Salamon}}, \bibinfo {author} {\bibfnamefont {A.}~\bibnamefont {Celi}},
  \bibinfo {author} {\bibfnamefont {R.~W.}\ \bibnamefont {Chhajlany}}, \bibinfo
  {author} {\bibfnamefont {I.}~\bibnamefont {Fr\'erot}}, \bibinfo {author}
  {\bibfnamefont {M.}~\bibnamefont {Lewenstein}}, \bibinfo {author}
  {\bibfnamefont {L.}~\bibnamefont {Tarruell}},\ and\ \bibinfo {author}
  {\bibfnamefont {D.}~\bibnamefont {Rakshit}},\ }\bibfield  {title} {\bibinfo
  {title} {Simulating twistronics without a twist},\ }\href
  {https://doi.org/10.1103/PhysRevLett.125.030504} {\bibfield  {journal}
  {\bibinfo  {journal} {Phys. Rev. Lett.}\ }\textbf {\bibinfo {volume} {125}},\
  \bibinfo {pages} {030504} (\bibinfo {year} {2020}{\natexlab{c}})}\BibitemShut
  {NoStop}%
\bibitem [{\citenamefont {Aidelsburger}\ \emph {et~al.}(2015)\citenamefont
  {Aidelsburger}, \citenamefont {Lohse}, \citenamefont {Schweizer},
  \citenamefont {Atala}, \citenamefont {Barreiro}, \citenamefont
  {Nascimb{\`e}ne}, \citenamefont {Cooper}, \citenamefont {Bloch},\ and\
  \citenamefont {Goldman}}]{aidelsburger2015measuring}%
  \BibitemOpen
  \bibfield  {author} {\bibinfo {author} {\bibfnamefont {M.}~\bibnamefont
  {Aidelsburger}}, \bibinfo {author} {\bibfnamefont {M.}~\bibnamefont {Lohse}},
  \bibinfo {author} {\bibfnamefont {C.}~\bibnamefont {Schweizer}}, \bibinfo
  {author} {\bibfnamefont {M.}~\bibnamefont {Atala}}, \bibinfo {author}
  {\bibfnamefont {J.~T.}\ \bibnamefont {Barreiro}}, \bibinfo {author}
  {\bibfnamefont {S.}~\bibnamefont {Nascimb{\`e}ne}}, \bibinfo {author}
  {\bibfnamefont {N.}~\bibnamefont {Cooper}}, \bibinfo {author} {\bibfnamefont
  {I.}~\bibnamefont {Bloch}},\ and\ \bibinfo {author} {\bibfnamefont
  {N.}~\bibnamefont {Goldman}},\ }\bibfield  {title} {\bibinfo {title}
  {Measuring the chern number of hofstadter bands with ultracold bosonic
  atoms},\ }\href@noop {} {\bibfield  {journal} {\bibinfo  {journal} {Nature
  Physics}\ }\textbf {\bibinfo {volume} {11}},\ \bibinfo {pages} {162}
  (\bibinfo {year} {2015})}\BibitemShut {NoStop}%
\bibitem [{\citenamefont {Huang}\ \emph {et~al.}(2016)\citenamefont {Huang},
  \citenamefont {Meng}, \citenamefont {Wang}, \citenamefont {Peng},
  \citenamefont {Zhang}, \citenamefont {Chen}, \citenamefont {Li},
  \citenamefont {Zhou},\ and\ \citenamefont {Zhang}}]{huang2016experimental}%
  \BibitemOpen
  \bibfield  {author} {\bibinfo {author} {\bibfnamefont {L.}~\bibnamefont
  {Huang}}, \bibinfo {author} {\bibfnamefont {Z.}~\bibnamefont {Meng}},
  \bibinfo {author} {\bibfnamefont {P.}~\bibnamefont {Wang}}, \bibinfo {author}
  {\bibfnamefont {P.}~\bibnamefont {Peng}}, \bibinfo {author} {\bibfnamefont
  {S.-L.}\ \bibnamefont {Zhang}}, \bibinfo {author} {\bibfnamefont
  {L.}~\bibnamefont {Chen}}, \bibinfo {author} {\bibfnamefont {D.}~\bibnamefont
  {Li}}, \bibinfo {author} {\bibfnamefont {Q.}~\bibnamefont {Zhou}},\ and\
  \bibinfo {author} {\bibfnamefont {J.}~\bibnamefont {Zhang}},\ }\bibfield
  {title} {\bibinfo {title} {Experimental realization of two-dimensional
  synthetic spin--orbit coupling in ultracold fermi gases},\ }\href@noop {}
  {\bibfield  {journal} {\bibinfo  {journal} {Nature Physics}\ }\textbf
  {\bibinfo {volume} {12}},\ \bibinfo {pages} {540} (\bibinfo {year}
  {2016})}\BibitemShut {NoStop}%
\bibitem [{\citenamefont {Wu}\ \emph {et~al.}(2016)\citenamefont {Wu},
  \citenamefont {Zhang}, \citenamefont {Sun}, \citenamefont {Xu}, \citenamefont
  {Wang}, \citenamefont {Ji}, \citenamefont {Deng}, \citenamefont {Chen},
  \citenamefont {Liu},\ and\ \citenamefont {Pan}}]{wu2016realization}%
  \BibitemOpen
  \bibfield  {author} {\bibinfo {author} {\bibfnamefont {Z.}~\bibnamefont
  {Wu}}, \bibinfo {author} {\bibfnamefont {L.}~\bibnamefont {Zhang}}, \bibinfo
  {author} {\bibfnamefont {W.}~\bibnamefont {Sun}}, \bibinfo {author}
  {\bibfnamefont {X.-T.}\ \bibnamefont {Xu}}, \bibinfo {author} {\bibfnamefont
  {B.-Z.}\ \bibnamefont {Wang}}, \bibinfo {author} {\bibfnamefont {S.-C.}\
  \bibnamefont {Ji}}, \bibinfo {author} {\bibfnamefont {Y.}~\bibnamefont
  {Deng}}, \bibinfo {author} {\bibfnamefont {S.}~\bibnamefont {Chen}}, \bibinfo
  {author} {\bibfnamefont {X.-J.}\ \bibnamefont {Liu}},\ and\ \bibinfo {author}
  {\bibfnamefont {J.-W.}\ \bibnamefont {Pan}},\ }\bibfield  {title} {\bibinfo
  {title} {Realization of two-dimensional spin-orbit coupling for bose-einstein
  condensates},\ }\href@noop {} {\bibfield  {journal} {\bibinfo  {journal}
  {Science}\ }\textbf {\bibinfo {volume} {354}},\ \bibinfo {pages} {83}
  (\bibinfo {year} {2016})}\BibitemShut {NoStop}%
\bibitem [{\citenamefont {Schreiber}\ \emph {et~al.}(2015)\citenamefont
  {Schreiber}, \citenamefont {Hodgman}, \citenamefont {Bordia}, \citenamefont
  {L{\"u}schen}, \citenamefont {Fischer}, \citenamefont {Vosk}, \citenamefont
  {Altman}, \citenamefont {Schneider},\ and\ \citenamefont
  {Bloch}}]{schreiber2015observation}%
  \BibitemOpen
  \bibfield  {author} {\bibinfo {author} {\bibfnamefont {M.}~\bibnamefont
  {Schreiber}}, \bibinfo {author} {\bibfnamefont {S.~S.}\ \bibnamefont
  {Hodgman}}, \bibinfo {author} {\bibfnamefont {P.}~\bibnamefont {Bordia}},
  \bibinfo {author} {\bibfnamefont {H.~P.}\ \bibnamefont {L{\"u}schen}},
  \bibinfo {author} {\bibfnamefont {M.~H.}\ \bibnamefont {Fischer}}, \bibinfo
  {author} {\bibfnamefont {R.}~\bibnamefont {Vosk}}, \bibinfo {author}
  {\bibfnamefont {E.}~\bibnamefont {Altman}}, \bibinfo {author} {\bibfnamefont
  {U.}~\bibnamefont {Schneider}},\ and\ \bibinfo {author} {\bibfnamefont
  {I.}~\bibnamefont {Bloch}},\ }\bibfield  {title} {\bibinfo {title}
  {Observation of many-body localization of interacting fermions in a
  quasirandom optical lattice},\ }\href@noop {} {\bibfield  {journal} {\bibinfo
   {journal} {Science}\ }\textbf {\bibinfo {volume} {349}},\ \bibinfo {pages}
  {842} (\bibinfo {year} {2015})}\BibitemShut {NoStop}%
\bibitem [{\citenamefont {Viebahn}\ \emph {et~al.}(2019)\citenamefont
  {Viebahn}, \citenamefont {Sbroscia}, \citenamefont {Carter}, \citenamefont
  {Yu},\ and\ \citenamefont {Schneider}}]{PhysRevLett.122.110404}%
  \BibitemOpen
  \bibfield  {author} {\bibinfo {author} {\bibfnamefont {K.}~\bibnamefont
  {Viebahn}}, \bibinfo {author} {\bibfnamefont {M.}~\bibnamefont {Sbroscia}},
  \bibinfo {author} {\bibfnamefont {E.}~\bibnamefont {Carter}}, \bibinfo
  {author} {\bibfnamefont {J.-C.}\ \bibnamefont {Yu}},\ and\ \bibinfo {author}
  {\bibfnamefont {U.}~\bibnamefont {Schneider}},\ }\bibfield  {title} {\bibinfo
  {title} {Matter-wave diffraction from a quasicrystalline optical lattice},\
  }\href {https://doi.org/10.1103/PhysRevLett.122.110404} {\bibfield  {journal}
  {\bibinfo  {journal} {Phys. Rev. Lett.}\ }\textbf {\bibinfo {volume} {122}},\
  \bibinfo {pages} {110404} (\bibinfo {year} {2019})}\BibitemShut {NoStop}%
\bibitem [{\citenamefont {Wang}\ \emph
  {et~al.}(2021{\natexlab{b}})\citenamefont {Wang}, \citenamefont {Cheng},
  \citenamefont {Wang}, \citenamefont {Zhang}, \citenamefont {Lu},
  \citenamefont {Yi}, \citenamefont {Niu}, \citenamefont {Deng}, \citenamefont
  {Liu}, \citenamefont {Chen} \emph {et~al.}}]{wang2021realization}%
  \BibitemOpen
  \bibfield  {author} {\bibinfo {author} {\bibfnamefont {Z.-Y.}\ \bibnamefont
  {Wang}}, \bibinfo {author} {\bibfnamefont {X.-C.}\ \bibnamefont {Cheng}},
  \bibinfo {author} {\bibfnamefont {B.-Z.}\ \bibnamefont {Wang}}, \bibinfo
  {author} {\bibfnamefont {J.-Y.}\ \bibnamefont {Zhang}}, \bibinfo {author}
  {\bibfnamefont {Y.-H.}\ \bibnamefont {Lu}}, \bibinfo {author} {\bibfnamefont
  {C.-R.}\ \bibnamefont {Yi}}, \bibinfo {author} {\bibfnamefont
  {S.}~\bibnamefont {Niu}}, \bibinfo {author} {\bibfnamefont {Y.}~\bibnamefont
  {Deng}}, \bibinfo {author} {\bibfnamefont {X.-J.}\ \bibnamefont {Liu}},
  \bibinfo {author} {\bibfnamefont {S.}~\bibnamefont {Chen}}, \emph {et~al.},\
  }\bibfield  {title} {\bibinfo {title} {Realization of an ideal weyl semimetal
  band in a quantum gas with 3d spin-orbit coupling},\ }\href@noop {}
  {\bibfield  {journal} {\bibinfo  {journal} {Science}\ }\textbf {\bibinfo
  {volume} {372}},\ \bibinfo {pages} {271} (\bibinfo {year}
  {2021}{\natexlab{b}})}\BibitemShut {NoStop}%
\bibitem [{\citenamefont {Foot}(2004)}]{foot2004atomic}%
  \BibitemOpen
  \bibfield  {author} {\bibinfo {author} {\bibfnamefont {C.~J.}\ \bibnamefont
  {Foot}},\ }\href@noop {} {\emph {\bibinfo {title} {Atomic physics}}},\
  Vol.~\bibinfo {volume} {7}\ (\bibinfo  {publisher} {OUP Oxford},\ \bibinfo
  {year} {2004})\BibitemShut {NoStop}%
\bibitem [{\citenamefont {Stewart}\ \emph {et~al.}(2008)\citenamefont
  {Stewart}, \citenamefont {Gaebler},\ and\ \citenamefont
  {Jin}}]{stewart2008using}%
  \BibitemOpen
  \bibfield  {author} {\bibinfo {author} {\bibfnamefont {J.}~\bibnamefont
  {Stewart}}, \bibinfo {author} {\bibfnamefont {J.}~\bibnamefont {Gaebler}},\
  and\ \bibinfo {author} {\bibfnamefont {D.}~\bibnamefont {Jin}},\ }\bibfield
  {title} {\bibinfo {title} {Using photoemission spectroscopy to probe a
  strongly interacting fermi gas},\ }\href@noop {} {\bibfield  {journal}
  {\bibinfo  {journal} {Nature}\ }\textbf {\bibinfo {volume} {454}},\ \bibinfo
  {pages} {744} (\bibinfo {year} {2008})}\BibitemShut {NoStop}%
\bibitem [{\citenamefont {Gon{\c{c}}alves}\ \emph {et~al.}(2021)\citenamefont
  {Gon{\c{c}}alves}, \citenamefont {Olyaei}, \citenamefont {Amorim},
  \citenamefont {Mondaini}, \citenamefont {Ribeiro},\ and\ \citenamefont
  {Castro}}]{gonccalves2021incommensurability}%
  \BibitemOpen
  \bibfield  {author} {\bibinfo {author} {\bibfnamefont {M.}~\bibnamefont
  {Gon{\c{c}}alves}}, \bibinfo {author} {\bibfnamefont {H.~Z.}\ \bibnamefont
  {Olyaei}}, \bibinfo {author} {\bibfnamefont {B.}~\bibnamefont {Amorim}},
  \bibinfo {author} {\bibfnamefont {R.}~\bibnamefont {Mondaini}}, \bibinfo
  {author} {\bibfnamefont {P.}~\bibnamefont {Ribeiro}},\ and\ \bibinfo {author}
  {\bibfnamefont {E.~V.}\ \bibnamefont {Castro}},\ }\bibfield  {title}
  {\bibinfo {title} {Incommensurability-induced sub-ballistic
  narrow-band-states in twisted bilayer graphene},\ }\href@noop {} {\bibfield
  {journal} {\bibinfo  {journal} {2D Materials}\ }\textbf {\bibinfo {volume}
  {9}},\ \bibinfo {pages} {011001} (\bibinfo {year} {2021})}\BibitemShut
  {NoStop}%
\bibitem [{\citenamefont {Lisi}\ \emph {et~al.}(2021)\citenamefont {Lisi},
  \citenamefont {Lu}, \citenamefont {Benschop}, \citenamefont {de~Jong},
  \citenamefont {Stepanov}, \citenamefont {Duran}, \citenamefont {Margot},
  \citenamefont {Cucchi}, \citenamefont {Cappelli}, \citenamefont {Hunter}
  \emph {et~al.}}]{lisi2021observation}%
  \BibitemOpen
  \bibfield  {author} {\bibinfo {author} {\bibfnamefont {S.}~\bibnamefont
  {Lisi}}, \bibinfo {author} {\bibfnamefont {X.}~\bibnamefont {Lu}}, \bibinfo
  {author} {\bibfnamefont {T.}~\bibnamefont {Benschop}}, \bibinfo {author}
  {\bibfnamefont {T.~A.}\ \bibnamefont {de~Jong}}, \bibinfo {author}
  {\bibfnamefont {P.}~\bibnamefont {Stepanov}}, \bibinfo {author}
  {\bibfnamefont {J.~R.}\ \bibnamefont {Duran}}, \bibinfo {author}
  {\bibfnamefont {F.}~\bibnamefont {Margot}}, \bibinfo {author} {\bibfnamefont
  {I.}~\bibnamefont {Cucchi}}, \bibinfo {author} {\bibfnamefont
  {E.}~\bibnamefont {Cappelli}}, \bibinfo {author} {\bibfnamefont
  {A.}~\bibnamefont {Hunter}}, \emph {et~al.},\ }\bibfield  {title} {\bibinfo
  {title} {Observation of flat bands in twisted bilayer graphene},\ }\href@noop
  {} {\bibfield  {journal} {\bibinfo  {journal} {Nature Physics}\ }\textbf
  {\bibinfo {volume} {17}},\ \bibinfo {pages} {189} (\bibinfo {year}
  {2021})}\BibitemShut {NoStop}%
\bibitem [{\citenamefont {Utama}\ \emph {et~al.}(2021)\citenamefont {Utama},
  \citenamefont {Koch}, \citenamefont {Lee}, \citenamefont {Leconte},
  \citenamefont {Li}, \citenamefont {Zhao}, \citenamefont {Jiang},
  \citenamefont {Zhu}, \citenamefont {Watanabe}, \citenamefont {Taniguchi}
  \emph {et~al.}}]{utama2021visualization}%
  \BibitemOpen
  \bibfield  {author} {\bibinfo {author} {\bibfnamefont {M.}~\bibnamefont
  {Utama}}, \bibinfo {author} {\bibfnamefont {R.~J.}\ \bibnamefont {Koch}},
  \bibinfo {author} {\bibfnamefont {K.}~\bibnamefont {Lee}}, \bibinfo {author}
  {\bibfnamefont {N.}~\bibnamefont {Leconte}}, \bibinfo {author} {\bibfnamefont
  {H.}~\bibnamefont {Li}}, \bibinfo {author} {\bibfnamefont {S.}~\bibnamefont
  {Zhao}}, \bibinfo {author} {\bibfnamefont {L.}~\bibnamefont {Jiang}},
  \bibinfo {author} {\bibfnamefont {J.}~\bibnamefont {Zhu}}, \bibinfo {author}
  {\bibfnamefont {K.}~\bibnamefont {Watanabe}}, \bibinfo {author}
  {\bibfnamefont {T.}~\bibnamefont {Taniguchi}}, \emph {et~al.},\ }\bibfield
  {title} {\bibinfo {title} {Visualization of the flat electronic band in
  twisted bilayer graphene near the magic angle twist},\ }\href@noop {}
  {\bibfield  {journal} {\bibinfo  {journal} {Nature Physics}\ }\textbf
  {\bibinfo {volume} {17}},\ \bibinfo {pages} {184} (\bibinfo {year}
  {2021})}\BibitemShut {NoStop}%
\bibitem [{\citenamefont {Wei\ss{}e}\ \emph {et~al.}(2006)\citenamefont
  {Wei\ss{}e}, \citenamefont {Wellein}, \citenamefont {Alvermann},\ and\
  \citenamefont {Fehske}}]{WeisseFehske2006}%
  \BibitemOpen
  \bibfield  {author} {\bibinfo {author} {\bibfnamefont {A.}~\bibnamefont
  {Wei\ss{}e}}, \bibinfo {author} {\bibfnamefont {G.}~\bibnamefont {Wellein}},
  \bibinfo {author} {\bibfnamefont {A.}~\bibnamefont {Alvermann}},\ and\
  \bibinfo {author} {\bibfnamefont {H.}~\bibnamefont {Fehske}},\ }\bibfield
  {title} {\bibinfo {title} {The kernel polynomial method},\ }\href
  {https://doi.org/10.1103/RevModPhys.78.275} {\bibfield  {journal} {\bibinfo
  {journal} {Rev. Mod. Phys.}\ }\textbf {\bibinfo {volume} {78}},\ \bibinfo
  {pages} {275} (\bibinfo {year} {2006})}\BibitemShut {NoStop}%
\bibitem [{\citenamefont {Pixley}\ \emph {et~al.}(2017)\citenamefont {Pixley},
  \citenamefont {Chou}, \citenamefont {Goswami}, \citenamefont {Huse},
  \citenamefont {Nandkishore}, \citenamefont {Radzihovsky},\ and\ \citenamefont
  {Das~Sarma}}]{PixleyDasSarma2017}%
  \BibitemOpen
  \bibfield  {author} {\bibinfo {author} {\bibfnamefont {J.~H.}\ \bibnamefont
  {Pixley}}, \bibinfo {author} {\bibfnamefont {Y.-Z.}\ \bibnamefont {Chou}},
  \bibinfo {author} {\bibfnamefont {P.}~\bibnamefont {Goswami}}, \bibinfo
  {author} {\bibfnamefont {D.~A.}\ \bibnamefont {Huse}}, \bibinfo {author}
  {\bibfnamefont {R.}~\bibnamefont {Nandkishore}}, \bibinfo {author}
  {\bibfnamefont {L.}~\bibnamefont {Radzihovsky}},\ and\ \bibinfo {author}
  {\bibfnamefont {S.}~\bibnamefont {Das~Sarma}},\ }\bibfield  {title} {\bibinfo
  {title} {Single-particle excitations in disordered weyl fluids},\ }\href
  {https://doi.org/10.1103/PhysRevB.95.235101} {\bibfield  {journal} {\bibinfo
  {journal} {Phys. Rev. B}\ }\textbf {\bibinfo {volume} {95}},\ \bibinfo
  {pages} {235101} (\bibinfo {year} {2017})}\BibitemShut {NoStop}%
\bibitem [{\citenamefont {Anderson}(1958)}]{Anderson1958}%
  \BibitemOpen
  \bibfield  {author} {\bibinfo {author} {\bibfnamefont {P.~W.}\ \bibnamefont
  {Anderson}},\ }\bibfield  {title} {\bibinfo {title} {Absence of diffusion in
  certain random lattices},\ }\href {https://doi.org/10.1103/PhysRev.109.1492}
  {\bibfield  {journal} {\bibinfo  {journal} {Phys. Rev.}\ }\textbf {\bibinfo
  {volume} {109}},\ \bibinfo {pages} {1492} (\bibinfo {year}
  {1958})}\BibitemShut {NoStop}%
\bibitem [{\citenamefont {Pixley}\ \emph {et~al.}(2016)\citenamefont {Pixley},
  \citenamefont {Huse},\ and\ \citenamefont {Das~Sarma}}]{PhysRevB.94.121107}%
  \BibitemOpen
  \bibfield  {author} {\bibinfo {author} {\bibfnamefont {J.~H.}\ \bibnamefont
  {Pixley}}, \bibinfo {author} {\bibfnamefont {D.~A.}\ \bibnamefont {Huse}},\
  and\ \bibinfo {author} {\bibfnamefont {S.}~\bibnamefont {Das~Sarma}},\
  }\bibfield  {title} {\bibinfo {title} {Uncovering the hidden quantum critical
  point in disordered massless dirac and weyl semimetals},\ }\href
  {https://doi.org/10.1103/PhysRevB.94.121107} {\bibfield  {journal} {\bibinfo
  {journal} {Phys. Rev. B}\ }\textbf {\bibinfo {volume} {94}},\ \bibinfo
  {pages} {121107} (\bibinfo {year} {2016})}\BibitemShut {NoStop}%
\bibitem [{\citenamefont {Gaunt}\ \emph {et~al.}(2013)\citenamefont {Gaunt},
  \citenamefont {Schmidutz}, \citenamefont {Gotlibovych}, \citenamefont
  {Smith},\ and\ \citenamefont {Hadzibabic}}]{PhysRevLett.110.200406}%
  \BibitemOpen
  \bibfield  {author} {\bibinfo {author} {\bibfnamefont {A.~L.}\ \bibnamefont
  {Gaunt}}, \bibinfo {author} {\bibfnamefont {T.~F.}\ \bibnamefont
  {Schmidutz}}, \bibinfo {author} {\bibfnamefont {I.}~\bibnamefont
  {Gotlibovych}}, \bibinfo {author} {\bibfnamefont {R.~P.}\ \bibnamefont
  {Smith}},\ and\ \bibinfo {author} {\bibfnamefont {Z.}~\bibnamefont
  {Hadzibabic}},\ }\bibfield  {title} {\bibinfo {title} {Bose-einstein
  condensation of atoms in a uniform potential},\ }\href
  {https://doi.org/10.1103/PhysRevLett.110.200406} {\bibfield  {journal}
  {\bibinfo  {journal} {Phys. Rev. Lett.}\ }\textbf {\bibinfo {volume} {110}},\
  \bibinfo {pages} {200406} (\bibinfo {year} {2013})}\BibitemShut {NoStop}%
\bibitem [{\citenamefont {Sokoloff}(1980)}]{Sokoloff1980}%
  \BibitemOpen
  \bibfield  {author} {\bibinfo {author} {\bibfnamefont {J.~B.}\ \bibnamefont
  {Sokoloff}},\ }\bibfield  {title} {\bibinfo {title} {Electron localization in
  crystals with quasiperiodic lattice potentials},\ }\href
  {https://doi.org/10.1103/PhysRevB.22.5823} {\bibfield  {journal} {\bibinfo
  {journal} {Phys. Rev. B}\ }\textbf {\bibinfo {volume} {22}},\ \bibinfo
  {pages} {5823} (\bibinfo {year} {1980})}\BibitemShut {NoStop}%
\bibitem [{\citenamefont {Zhu}\ \emph {et~al.}(2021)\citenamefont {Zhu},
  \citenamefont {Shi},\ and\ \citenamefont {MacDonald}}]{PhysRevB.103.235146}%
  \BibitemOpen
  \bibfield  {author} {\bibinfo {author} {\bibfnamefont {J.}~\bibnamefont
  {Zhu}}, \bibinfo {author} {\bibfnamefont {J.}~\bibnamefont {Shi}},\ and\
  \bibinfo {author} {\bibfnamefont {A.~H.}\ \bibnamefont {MacDonald}},\
  }\bibfield  {title} {\bibinfo {title} {Theory of angle-resolved photoemission
  spectroscopy in graphene-based moir\'e superlattices},\ }\href
  {https://doi.org/10.1103/PhysRevB.103.235146} {\bibfield  {journal} {\bibinfo
   {journal} {Phys. Rev. B}\ }\textbf {\bibinfo {volume} {103}},\ \bibinfo
  {pages} {235146} (\bibinfo {year} {2021})}\BibitemShut {NoStop}%
\bibitem [{\citenamefont {Ahn}\ \emph {et~al.}(2018)\citenamefont {Ahn},
  \citenamefont {Moon}, \citenamefont {Kim}, \citenamefont {Kim}, \citenamefont
  {Shin}, \citenamefont {Kim}, \citenamefont {Cha}, \citenamefont {Kahng},
  \citenamefont {Kim}, \citenamefont {Koshino} \emph {et~al.}}]{Ahn2018}%
  \BibitemOpen
  \bibfield  {author} {\bibinfo {author} {\bibfnamefont {S.~J.}\ \bibnamefont
  {Ahn}}, \bibinfo {author} {\bibfnamefont {P.}~\bibnamefont {Moon}}, \bibinfo
  {author} {\bibfnamefont {T.-H.}\ \bibnamefont {Kim}}, \bibinfo {author}
  {\bibfnamefont {H.-W.}\ \bibnamefont {Kim}}, \bibinfo {author} {\bibfnamefont
  {H.-C.}\ \bibnamefont {Shin}}, \bibinfo {author} {\bibfnamefont {E.~H.}\
  \bibnamefont {Kim}}, \bibinfo {author} {\bibfnamefont {H.~W.}\ \bibnamefont
  {Cha}}, \bibinfo {author} {\bibfnamefont {S.-J.}\ \bibnamefont {Kahng}},
  \bibinfo {author} {\bibfnamefont {P.}~\bibnamefont {Kim}}, \bibinfo {author}
  {\bibfnamefont {M.}~\bibnamefont {Koshino}}, \emph {et~al.},\ }\bibfield
  {title} {\bibinfo {title} {Dirac electrons in a dodecagonal graphene
  quasicrystal},\ }\href@noop {} {\bibfield  {journal} {\bibinfo  {journal}
  {Science}\ }\textbf {\bibinfo {volume} {361}},\ \bibinfo {pages} {782}
  (\bibinfo {year} {2018})}\BibitemShut {NoStop}%
\bibitem [{\citenamefont {Park}\ \emph {et~al.}(2019)\citenamefont {Park},
  \citenamefont {Kim},\ and\ \citenamefont {Lee}}]{ParkLee2019}%
  \BibitemOpen
  \bibfield  {author} {\bibinfo {author} {\bibfnamefont {M.~J.}\ \bibnamefont
  {Park}}, \bibinfo {author} {\bibfnamefont {H.~S.}\ \bibnamefont {Kim}},\ and\
  \bibinfo {author} {\bibfnamefont {S.}~\bibnamefont {Lee}},\ }\bibfield
  {title} {\bibinfo {title} {Emergent localization in dodecagonal bilayer
  quasicrystals},\ }\href {https://doi.org/10.1103/PhysRevB.99.245401}
  {\bibfield  {journal} {\bibinfo  {journal} {Phys. Rev. B}\ }\textbf {\bibinfo
  {volume} {99}},\ \bibinfo {pages} {245401} (\bibinfo {year}
  {2019})}\BibitemShut {NoStop}%
\bibitem [{\citenamefont {Attig}\ \emph {et~al.}(2021)\citenamefont {Attig},
  \citenamefont {Park}, \citenamefont {Scherer}, \citenamefont {Trebst},
  \citenamefont {Altland},\ and\ \citenamefont {Rosch}}]{AttigRosch2021}%
  \BibitemOpen
  \bibfield  {author} {\bibinfo {author} {\bibfnamefont {J.}~\bibnamefont
  {Attig}}, \bibinfo {author} {\bibfnamefont {J.}~\bibnamefont {Park}},
  \bibinfo {author} {\bibfnamefont {M.~M.}\ \bibnamefont {Scherer}}, \bibinfo
  {author} {\bibfnamefont {S.}~\bibnamefont {Trebst}}, \bibinfo {author}
  {\bibfnamefont {A.}~\bibnamefont {Altland}},\ and\ \bibinfo {author}
  {\bibfnamefont {A.}~\bibnamefont {Rosch}},\ }\bibfield  {title} {\bibinfo
  {title} {Universal principles of moir{\'{e}} band structures},\ }\href
  {https://doi.org/10.1088/2053-1583/ac1cf0} {\bibfield  {journal} {\bibinfo
  {journal} {2D Materials}\ }\textbf {\bibinfo {volume} {8}},\ \bibinfo {pages}
  {044007} (\bibinfo {year} {2021})}\BibitemShut {NoStop}%
\bibitem [{Aub(1980)}]{AubryAndre1980}%
  \BibitemOpen
  \href@noop {} {\bibfield  {journal} {\bibinfo  {journal} {Ann. Isr. Phys.
  Soc.}\ }\textbf {\bibinfo {volume} {3}} (\bibinfo {year} {1980})}\BibitemShut
  {NoStop}%
\bibitem [{\citenamefont {Devakul}\ and\ \citenamefont
  {Huse}(2017)}]{DevakulHuse2017}%
  \BibitemOpen
  \bibfield  {author} {\bibinfo {author} {\bibfnamefont {T.}~\bibnamefont
  {Devakul}}\ and\ \bibinfo {author} {\bibfnamefont {D.~A.}\ \bibnamefont
  {Huse}},\ }\bibfield  {title} {\bibinfo {title} {Anderson localization
  transitions with and without random potentials},\ }\href
  {https://doi.org/10.1103/PhysRevB.96.214201} {\bibfield  {journal} {\bibinfo
  {journal} {Phys. Rev. B}\ }\textbf {\bibinfo {volume} {96}},\ \bibinfo
  {pages} {214201} (\bibinfo {year} {2017})}\BibitemShut {NoStop}%
\bibitem [{\citenamefont {Szab\'o}\ and\ \citenamefont
  {Schneider}(2020)}]{SzaboSchndeider2020}%
  \BibitemOpen
  \bibfield  {author} {\bibinfo {author} {\bibfnamefont {A.}~\bibnamefont
  {Szab\'o}}\ and\ \bibinfo {author} {\bibfnamefont {U.}~\bibnamefont
  {Schneider}},\ }\bibfield  {title} {\bibinfo {title} {Mixed spectra and
  partially extended states in a two-dimensional quasiperiodic model},\ }\href
  {https://doi.org/10.1103/PhysRevB.101.014205} {\bibfield  {journal} {\bibinfo
   {journal} {Phys. Rev. B}\ }\textbf {\bibinfo {volume} {101}},\ \bibinfo
  {pages} {014205} (\bibinfo {year} {2020})}\BibitemShut {NoStop}%
\bibitem [{ama()}]{amarel}%
  \BibitemOpen
  \href@noop {} {}\bibinfo {howpublished}
  {\url{https://oarc.rutgers.edu/}}\BibitemShut {NoStop}%
\bibitem [{\citenamefont {Mott}(1967)}]{Mott1967}%
  \BibitemOpen
  \bibfield  {author} {\bibinfo {author} {\bibfnamefont {N.}~\bibnamefont
  {Mott}},\ }\bibfield  {title} {\bibinfo {title} {Electrons in disordered
  structures},\ }\href@noop {} {\bibfield  {journal} {\bibinfo  {journal}
  {Advances in Physics}\ }\textbf {\bibinfo {volume} {16}},\ \bibinfo {pages}
  {49} (\bibinfo {year} {1967})}\BibitemShut {NoStop}%
\end{thebibliography}%

\end{document}